\documentclass[twocolumn]{aastex61}
\usepackage{amsmath,float,xcolor}
\usepackage{natbib} \usepackage{epstopdf} 
 \graphicspath{{./pic/}}

\newcommand{\angstrom}{\text{\normalfont\AA}}
\newcommand{\RN}[1]{\textup{\uppercase\expandafter{\romannumeral#1}}}
\shorttitle{Stellar Metallicity} \shortauthors{Leethochawalit et al.}

\begin{document}
\title{Evolution of the Stellar Mass--Metallicity Relation - I: Galaxies in the z $\sim0.4$ Cluster Cl0024} 
\correspondingauthor{Nicha Leethochawalit}
\author{Nicha Leethochawalit} \affil{Cahill Center for Astronomy and
Astrophysics, California Institute of Technology, MS 249-17, Pasadena,
CA 91125}
\author{Evan N. Kirby} \affil{Cahill Center for Astronomy and
Astrophysics, California Institute of Technology, MS 249-17, Pasadena,
CA 91125}
\author{Sean M. Moran} \affil{Smithsonian Astrophysical Observatory, 
60 Garden St, Cambridge, MA 02138}
\author{Richard S. Ellis} \affil{Department of Physics and Astronomy, 
University College London, Gower Street, London WC1E 6BT, UK}
\author{Tommaso Treu} \affil{Department of Physics and Astronomy, 
University of California, Los Angeles, CA 90095}

\begin{abstract}
We present the stellar mass--stellar metallicity relationship (MZR) in
the Cl0024+1654 galaxy cluster at $z\sim0.4$ using full spectrum stellar
population synthesis modeling of individual quiescent galaxies. The
lower limit of our stellar mass range is $M_*=10^{9.7}M_\odot$,
the lowest galaxy mass at which individual stellar metallicity has been measured 
beyond the local universe. We report a detection of an 
evolution of the stellar MZR with observed redshift at $0.037\pm0.007$ dex per Gyr, 
consistent with the predictions from hydrodynamical simulations. Additionally, 
we find that the evolution of the stellar MZR with observed redshift 
can be explained by an evolution of the stellar MZR with their formation time, 
i.e., when the single stellar population (SSP)-
equivalent ages of galaxies are taken into account. 
This behavior is consistent with 
stars forming out of gas that also has an MZR with a normalization 
that decreases with redshift. Lastly, we find that over the observed 
mass range, the MZR can be described by a linear function with a 
shallow slope, (${\rm [Fe/H]} \propto {(0.16 \pm 0.03)} \log M_*$). The slope 
suggests that galaxy feedback, in terms of mass-loading factor, might 
be mass-independent over the observed mass and redshift range.
\end{abstract}


\section{Introduction}\label{sec:intro}
One of the best-known properties in observed galaxies 
is the tight correlation between galaxy stellar mass 
and gas-phase metallicity, i.e., the mass--metallicity 
relation (MZR). Several large galaxy surveys, such as 
the Sloan Digital Sky Survey (SDSS), have confirmed 
that galaxies at all redshifts with higher stellar masses
retain more metals than galaxies with lower stellar masses
\citep[e.g.][]{Tremonti04,Sanders15,Guo16,Onodera16}. 
While the details of the evolution of the gas-phase 
MZR over redshift are still debated, mainly due to the 
different metallicity indicators \citep[e.g.][]{Steidel14,Kewley15,Strom17,Bian17}, 
galaxies at higher redshifts generally follow the same 
trend as the local MZR but are somewhat offset toward 
lower metallicities \citep[e.g.,][]{Erb06,Maiolino08,Steidel14,Zahid14}.

Despite the well-established gas-phase MZR, 
our understanding of the amount of metals that stars 
incorporate from the gas is less secure. The gas-phase 
metallicity only indicates the amount of metals 
in the gas during the time of observation. The metallicity 
in stars indicates the metal content in stars at their 
formation. Therefore, measuring stellar metallicity from 
a galaxy's integrated stellar light can reveal the 
``star formation history-averaged'' galactic metallicity. It 
is less susceptible than the gas to instantaneous fluctuations. 
Obtaining stellar metallicities over a range of galaxy masses,
 i.e., the stellar MZR, can provide insight on the chemical 
evolution of galaxies complementary to the gas-phase MZR\@. 
For example, \citet{Peng15} compared the stellar MZR 
of local star-forming galaxies to that of quenched galaxies to 
study galaxy quenching mechanisms.

A few suites of cosmological hydrodynamical
simulations have made quantitative predictions of the stellar 
MZR in the past few years. \citet{Ma16} investigated the 
evolution of both gas-phase and stellar MZRs over redshift 
from a limited number of galaxies in the Feedback in Realistic 
Environment (FIRE) cosmological zoom-in simulations. 
These simulations suggest that the stellar MZR evolves
monotonically, with an increase in stellar metallicity of $\sim0.3$ 
dex at fixed stellar mass from $z=1$ to $z=0$. \citet{DeRossi17} 
presented stellar MZRs at four redshifts derived from a different 
suite of cosmological hydrodynamical simulations, the Evolution 
and Assembly of GaLaxies and their Environments (EAGLE). 
The derived stellar MZRs came from a larger number of
galaxies than those in \citeauthor{Ma16}'s (\citeyear{Ma16}) 
study but with coarser spatial resolution. \citet{DeRossi17} predicted 
that the evolution of the stellar MZR is 0.2 dex from $z=1$ 
to $z=0$ at a stellar mass of $10^{9.5}M_\odot$,
slightly smaller than predicted by \citet{Ma16}.

The classical approach for measuring ages and metallicities of stellar
populations is to use spectrophotometric indices such as Lick
indices \citep{Faber73, Worthey94}, where the equivalent widths of some
spectral features expected to correlate with metal abundance or age
are measured and compared to model predictions. \citet{Gallazzi05} 
conducted one of the pioneering works using spectrophotometric 
indices to measure the stellar MZR from local $z\sim0$ SDSS 
star-forming and quiescent galaxies. \citet{Gallazzi14} extended 
their work to 77 galaxies at $z\sim0.7$. They found an offset of 
the stellar MZR by 0.12 dex from $z=0.7$ to $z=0$ among the 
star-forming population but no significant evolution among the 
quiescent population. 

However, the stellar metallicities of star-forming galaxies should be
taken with caution since they tend to be more difficult to measure. 
Large uncertainties and biases could arise from emission line
subtractions, a lack of young stars in stellar libraries, and the fact that the
most luminous stars are not necessarily the majority of the mass. The
uncertainties of metallicities in star-forming galaxies measured by
\citet{Gallazzi06} are generally twice the uncertainties of the
metallicities in quiescent galaxies, with a median of $\delta \log Z
\sim 0.16$ for star forming galaxies and 0.08 for quiescent
galaxies. Moreover, using a similar sample of SDSS galaxies,
\citet{Panter08} found that star-forming galaxies have higher
stellar metallicity than the whole sample while
\citet{Gallazzi05} and \citet{Peng15} found the opposite. Adding further
confusion, when metallicities were measured from equivalent widths of
UV absorption lines, \citet{Sommariva12} found that the stellar MZR of
star-forming galaxies at $z=3$ is consistent with that of local galaxies
measured by \citet{Gallazzi05}, i.e., no significant evolution of the stellar MZR
among the star-forming galaxies from $z=3$ to $z=0$. This contradicts the
conclusion of \citet{Gallazzi14}. Nonetheless, the metallicities measured by
\citet{Sommariva12}---using light mainly produced by O stars---might trace 
different populations from those measured by
\citet{Gallazzi05}---using the light from stars of earlier spectral type.

If we focus on the quiescent galaxies whose ages and metallicities can be measured
more reliably, a number of recent works have been employing an
alternative approach---a full spectrum fitting technique---to
determine the ages and metallicities of these quiescent stellar populations. The
modeling of full optical--NIR spectra of stellar systems has been
advanced in the past decade \citep[e.g.,][]{CidFernandes05,
Ocvirk06, Walcher09, Conroy09, Conroy12, Conroy14}. The method
is preferred over the use of spectrophotometric indices because it
utilizes nearly all of the information from the collected light, resulting in smaller
uncertainties.

Though the approach has been used to measure ages and metallicities of
both local and high redshift galaxies, the measurements of higher
redshift galaxies are still mainly limited to stacks of spectra due to
the lack of sufficient signal in individual
spectra. \citet{Choi14} measured the stellar MZR from stacked spectra of
quiescent galaxies ranging from $z=0$ to $z=0.7$. The stellar MZRs
show possible evidence for evolution with
redshift. The MZR measured from stacked spectra can only reflect the
median metallicities of the population. It cannot reveal the the scatter of age or metallicity
within the population. In fact, \citet{Choi14} measured very different ages and
metallicities of two individual galaxies at $z=0.8$. The ages and
metallicities of the individual galaxies were significantly different from the results from the
stacked spectrum at the same redshift. Although the two galaxies 
were brightest cluster galaxies, the results suggest a possible large
scatter within the population. Moreover, measuring galaxy properties
in individual galaxies can reveal potentially important correlations
between galaxy parameters such as $\Delta$[Fe/H] and age or
$\Delta$[Fe/H] and [$\alpha$/Fe], where $\Delta$[Fe/H] is the deviation 
in an individual galaxy's metallicity from the MZR of the whole population.

Ultimately, the observations so far reveal no strong evidence for the
evolution of the stellar MZR, though the simulations suggest otherwise. 
If evolution is present, conceivably it has not been detected because 
earlier observations have been limited to the most massive galaxies.
\citet{DeRossi17} predicted that the evolution of the stellar
MZR is 0.2 dex from $z=1$ to $z=0$ for a mass of $10^{9.5}M_\odot$ but
less than $0.05$ dex at $10^{10.5}M_\odot$ over the same
redshift span. \citeauthor{Choi14}'s (\citeyear{Choi14}) sample at $z>0.4$ is limited to
massive galaxies ($M_*>10^{10.8}M_\odot$). The passive galaxies at $z=0.7$
in \citeauthor{Gallazzi14}'s (\citeyear{Gallazzi14}) sample are also massive
($M_*>10^{10.5}M_\odot$) and have large uncertainties in
[Fe/H] on the order of $\sim0.2$ dex.
\citet{Onodera15} used Lick indices to measure the age and
stellar metallicity of a stacked spectrum of passive galaxies
at $z = 1.6$, the highest redshift at which stellar metallicity has been measured. 
The masses were also limited to
$M_*>10^{11} M_\odot$, where minimal evolution is expected. Since all
of the observations at higher redshifts so far are limited to the
massive end, it is not surprising that there is no statistically significant evidence for the
evolution of the stellar MZR\@. In fact, the MZR of \citet{Gallazzi14} tentatively suggests a
stronger chemical evolution in lower mass galaxies even among the
quiescent populations. 

This paper is the first in a series to present individual stellar
metallicity measurements from $z=0$ to $z=1$. For the first time, we report a detection 
of $>5\sigma$ in the evolution of the stellar MZR with redshift based on the stellar MZR of 
62 early-type galaxies from the galaxy cluster Cl0024+1654 at $z\sim0.4$.
To our knowledge, this is the first attempt to measure
metallicities using full spectrum synthesis from individual spectra
that extend to stellar masses as low as $10^{9.7} M_\odot$ beyond the local universe. 
The major advances in this paper that made the detection possible are 
(a) the measurement of ages and metallicities
in individual galaxies at $z>0$; (b) the extension of the mass range to low-mass
galaxies; and (c) the use of full spectrum synthesis in deriving the
parameters. Aside from the evolution with observed redshift 
(at $0.037\pm0.007$ dex per Gyr), we also detect an even stronger 
evolution of the stellar MZR with the redshifts at which the galaxies formed 
(at $0.055\pm0.006$ dex per Gyr).

In Section \ref{sec:data}, we
describe the data and their completeness. In Section \ref{sec:model},
we describe the method we used to measure ages and metallicities
including how it performed as a function of signal-to-noise ratio and
when the assumption of the single stellar population was dropped. 
Further comparisons between our measurements and the measurements from 
the literature can be found in Appendix \ref{sec:compare_measurement}. In Section
\ref{sec:localmzr}, we present the MZR derived from a subsample of local
SDSS quiescent galaxies. In Section \ref{sec:main_results}, we
discuss our results and demonstrate that there is evolution in the stellar MZR
with both observed redshift and with formation redshift, i.e., when 
galaxy ages are taken into account. We also examine the relatively gentle
slope found in the observed MZR and how it relates to the feedback strength 
of galaxies in the observed mass range. Lastly, we discuss the impact of 
cluster environments on our results in Section \ref{sec:environment}. 
Throughout this paper, we assume cosmological parameters $\Omega_\Lambda = 0.7$, $\Omega_m = 0.3$ and 
$H_0=70 \textrm{ km s}^{-1} \textrm{Mpc}^{-1}$.

%%%%%%%%%%%%%%%%%%%%%%%%%%%%%%%%%%%%%%%%%%%%%%%%%%%%%%%%%%
%%%%%%%% SECTION 2 Data %%%%%%%%%%%%%%%%%%%%%%%%%%%%%%%%%%%%%%%%
%%%%%%%%%%%%%%%%%%%%%%%%%%%%%%%%%%%%%%%%%%%%%%%%%%%%%%%%%%
\section{Data}
\label{sec:data}
We leverage a large imaging and spectroscopic survey of the $z\simeq0.4$
galaxy cluster Cl0024+1654 \citep{Treu03, Moran05,Moran06, Moran07a,
Moran07b}. The study provides comprehensive Keck spectroscopy 
and Hubble Space Telescope (HST) imaging of Cl0024+17 members. 
Studying galaxy clusters offers the advantage of being able to obtain 
a large number of spectra in a few Keck pointings. 

\citet{Moran07a} provide complete details of the survey. In summary,
the HST imaging of Cl0024+17 consists of 39 sparsely sampled WFPC2
images taken in the F814W filter at an exposure time of 4-4.4 ks
each. The imaged field spans up to $>5$ Mpc from the cluster center, a
significantly larger radius than the cluster's virial radius of 1.7
Mpc. Supplementing the HST imaging, the study also provides infrared 
imaging in the $K_s$-band and $J$-band from the WIRC camera 
on the Hale 200-inch telescope, optical imaging in the $BVRI$ bands 
from the 3.6 m Canada-France-Hawaii Telescope, and both 
near- (NUV) and far-ultraviolet (FUV) imaging on the Galaxy 
Evolution Explorer (GALEX) satellite. We used the photometric catalog
from the survey's Web
site\footnote{\url{http://www.astro.caltech.edu/clusters/}} which
includes all optical and infrared photometric measurements, HST morphologies, and
photometric/spectral redshifts. We downloaded the UV catalog separately 
through the Mikulski Archive for Space Telescopes (MAST). All UV galaxy images 
were visually examined with the optical counterparts to ensure the 
correct associations.

The survey used DEIMOS \citep{Faber03} on the
Keck \RN{2} telescope to obtain deep spectra of 300 member 
galaxies to $M_V=-18$. The observations took place between 
2003 and 2005. The targets were selected with priority given to 
known cluster members up to $I=22.5$ with classified HST 
morphologies. A total of 16 masks were observed with integration 
times of 2-4 hrs each. Twelve masks were observed with the 
900 line mm$^{-1}$ grating from 2003 to 2004 while the rest 
were observed with the 600 line mm$^{-1}$ grating in 2005 
\citep{Moran07a}. All slitlets were $1\arcsec$ wide, yielding 
spectral resolutions of $R\sim2000-3000$. All masks were 
centered at $\sim6200$ \angstrom, providing a spectral range 
of 3500 to 5500 \angstrom\ in the rest frame. The spectroscopic 
sample is $>65\%$ complete for objects with $m_{\rm F814W}<21.1$. 
 
The DEIMOS spectra were reduced using the spec2d DEIMOS
data reduction pipeline \citep{Cooper12, Newman13} adapted 
by \citet{Kirby15}. Each spectrum was flat-fielded, wavelength-calibrated, 
sky-subtracted and telluric-corrected. 

\begin{figure}
	\centering
	\includegraphics[width=0.4\textwidth]{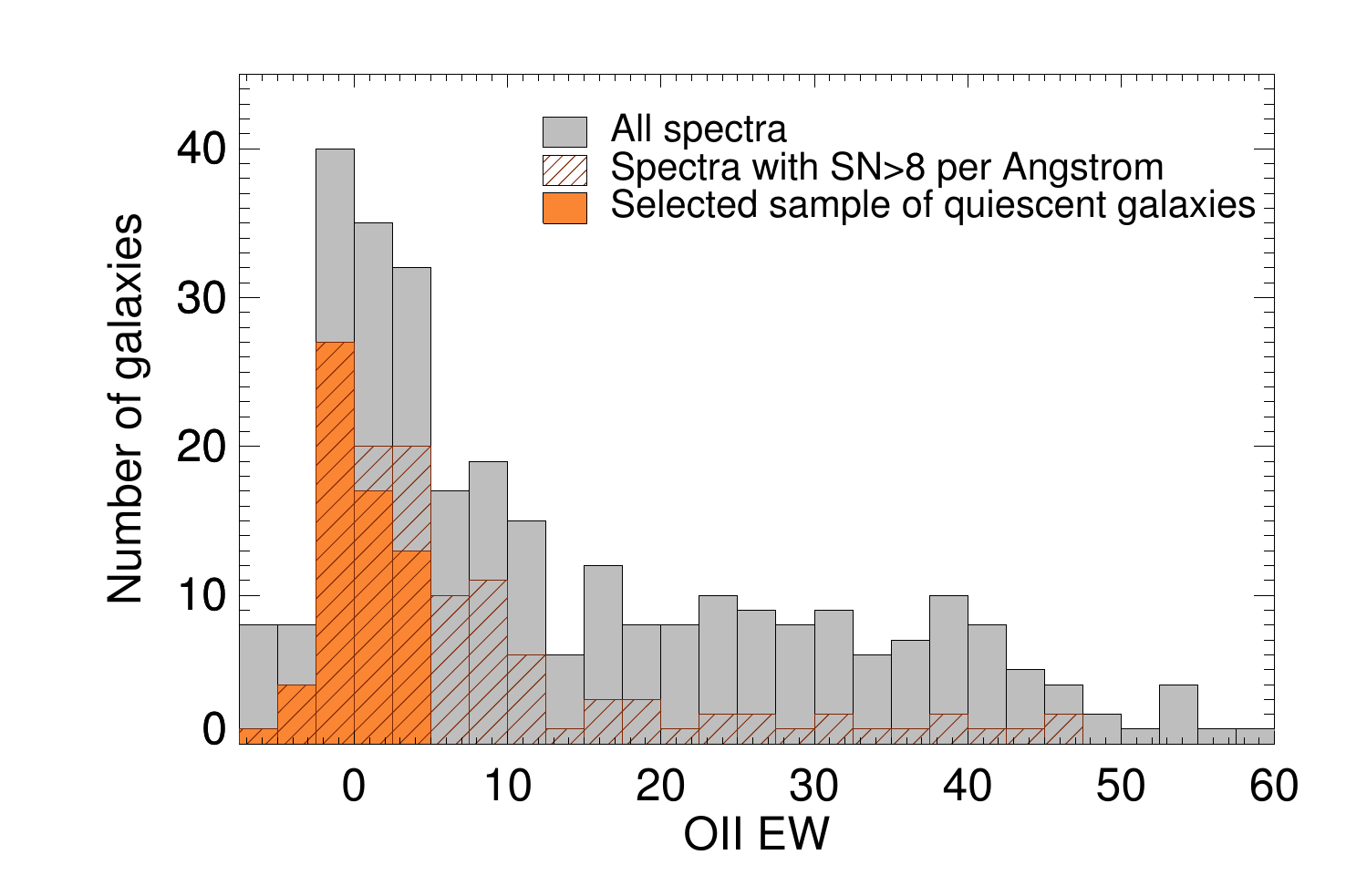}
	\includegraphics[width=0.4\textwidth]{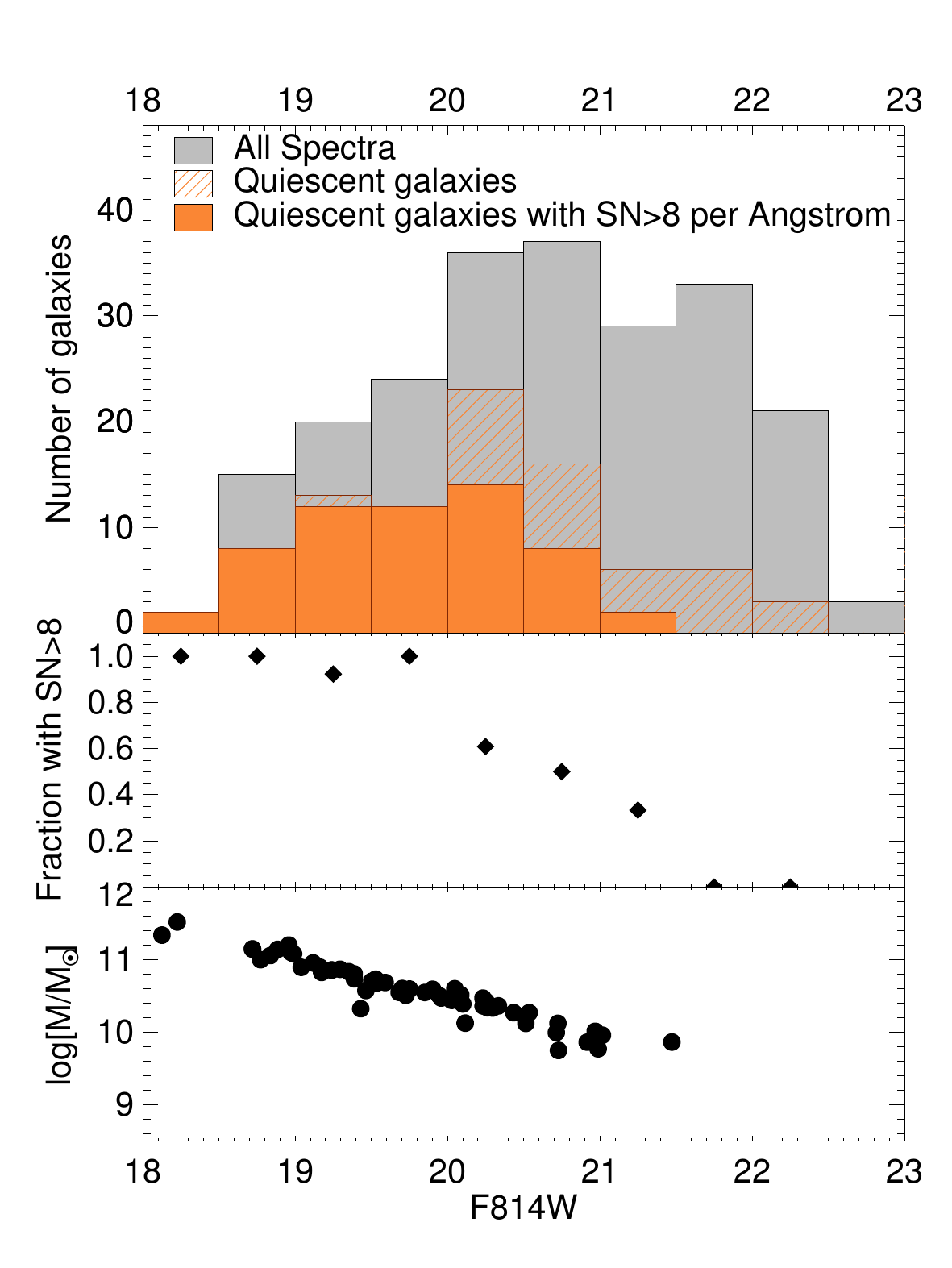}
	\caption{Characteristics of the observed spectra in the
Cl0024+17 cluster. ({\it Top}) Histograms of [\ion{O}{2}] $\lambda$3727
rest-frame equivalent widths. In this study, we use samples of
quiescent galaxies (EW $<5$ \angstrom\ and FUV$-$V $> 3$) with S/N $>8$
\angstrom$^{-1}$. ({\it Bottom}) Histograms of galaxies as a
function of F814W magnitude. The bottom panel shows magnitude
as a function of stellar mass. Our sample is $\sim50\%$
complete at $M_*>10^{9.7}M_\odot$. }
	\label{fig:data_charac}
\end{figure}

We selected subsamples of quiescent galaxies with an
average signal-to-noise ratio (S/N) greater than 8~\angstrom$^{-1}$ 
in the observed frame to ensure that we can break the degeneracy 
between age and metallicity (see Section \ref{sec:accuracy}). 
We define quiescent galaxies as those with rest-frame equivalent widths 
(EWs) of [\ion{O}{2}] $\lambda$3727 smaller than $5$ \angstrom\ 
and rest-frame FUV$-$V colors larger than 3. 
The EW limit is roughly equivalent to having a cut in specific 
star formation rate (sSFR)\footnote{The sSFR approximation 
is based on the SFR([\ion{O}{2}]) calibration from \citet{Kewley04}, 
the stellar mass-to-light (M/$L_B$) ratios from \citet{Bell01} and the mean 
rest-frame (U-B) color $\sim1.5$ of the sample.} at approximately 
$10^{-11}M_\odot$ yr$^{-1}$. The color cut is intended to further 
minimize contamination by star-forming galaxies. Specifically, the cut 
eliminates galaxies with star formation more recent than $10^7-10^8$ yr 
\citep[see][]{Moran06}. 
The spectral S/N is estimated as the inverse of the average 
ratio between the absolute deviation
of the observed spectrum from the best-fit spectrum of all
pixels in continuum region. We quote the S/N per \AA\ rather 
than per pixel.

The [\ion{O}{2}] EW, FUV$-$V color, and S/N criteria reduced the sample 
from the original total of 300 observed DEIMOS spectra to 
62 quiescent spectra with sufficient S/N\@. 
Figure \ref{fig:data_charac} shows the fraction of the
selected sample to the parent sample as a function of 
F814W magnitude and stellar mass. Stellar masses and rest-frame colors were
derived from available photometry using the SDSS KCORRECT software version 
v4\_3  \citep{Blanton07}, which assumes \citet{BC03} population synthesis models 
and the \citet{Chabrier03} stellar initial mass function. 
The fraction of quiescent galaxies with S/N $>8$ \angstrom$^{-1}$ is
$\sim75\%$ for F814W$<21.1$. Since the parent survey is $>65\%$ 
complete at the same magnitude range for the DEIMOS spectra, 
our final sample is therefore $\sim50\%$ complete for 
F814W$<21.1$ or $M_*\gtrsim10^{9.7}M_\odot$.

%%%%%%%%%%%%%%%%%%%%%%%%%%%%%%%%%%%%%%%%%%%%%%%%%%%%%%%%%%%%%
%%%%%%%% SECTION 3 Method %%%%%%%%%%%%%%%%%%%%%%%%%%%%%%%%%%%%%%%%%%%%
%%%%%%%%%%%%%%%%%%%%%%%%%%%%%%%%%%%%%%%%%%%%%%%%%%%%%%%%%%%%%%%
\section{Model Fitting}
\label{sec:model}
In this section, we describe the full-spectrum fitting technique used
to derive ages and metallicities. Section \ref{sec:general_model}
describes the stellar population synthesis models used in this
work. In Section \ref{sec:measurement}, we detail our fitting
technique. We test the accuracy of our measurements against
our assumptions in Section \ref{sec:accuracy}.

\subsection{Model Spectra}
\label{sec:general_model}
We derived stellar metallicities and ages of the galaxies using the
full-spectrum fitting technique via stellar population synthesis
(SPS). In the past decade, SPS models have been refined and
often used as alternative tools to photometric indices to derive
stellar population parameters such as stellar mass, star formation
history, metallicity, and age
\citep[e.g.,][]{Schiavon06,Walcher09,Choi14,Fumagalli16}. The commonly
used SPS models include \citet[][BC03]{BC03},
\citet[][PEGASE]{Fioc99}, \citet[][M05]{Maraston05}, and
\citet[][FSPS]{Conroy09}. The main advantage of using SPS over
photometric indices is that it utilizes information from the whole
spectrum simultaneously instead of using portions with the strongest
stellar absorption features. Consequently, spectra with lower
signal-to-noise ratios can be used to achieve the same level of
precision.

We adopted the Flexible Stellar Population Synthesis (FSPS) models
\citep{Conroy09} among the publicly available SPS models to derive the
stellar population parameters from our sample. FSPS utilizes the most
recent model of the Padova stellar evolution tracks
\citep{Marigo08}. The model includes treatments of thermally pulsating
asymptotic giant branch (TP-AGBs) and an option to include
horizontal-branch (HB) and blue straggler (BS) stars. The TP-AGBs are
particularly important because leaving them out can result in
systematic differences in age by a factor of 2 \citep{Maraston05,
Conroy09}. Lastly, the model offers flexibility in modeling the
spectra that match the spectral resolution and the ranges of stellar 
parameters in our study.

We generated the templates of single stellar population (SSP) spectra
using the FSPS code version 3.0. The SSP spectra were generated with
the Kroupa IMF \citep{Kroupa01} and the MILES spectral library
\citep{Sanchez-Blazques06}. We chose the MILES spectral library
because it spans a wide range in the stellar parameters, i.e., $\log g$
and [Fe/H], that are suitable for measuring galaxies at higher
redshifts with lower metallicity. In addition, the library
has a spectral resolution of 2.3 \angstrom\ FWHM across the wavelength
range of 3525 to 7500 \angstrom, comparable to that of our
spectra. The spectra were interpolated from 22 modeled metallicities 
ranging from $\log Z=-1.98$ to 0.2, corresponding to the metallicity 
values of the Padova isochrones. The age ranges from 0.3 Myr to 14 Gyr. 
The rest of the parameters were set to the default mode in generating SSP spectra,
which means dust, blue HB, and BS stars were excluded. Because dust
absorption mainly affects the continuum of the spectra but not the
absorption lines, and because we remove the continuum from the observed spectra
in our fitting procedure, we omitted dust for simplicity. The
omission of blue HBs and BS stars should not affect the derived
ages and metallicities because these stars mainly contribute to
ultraviolet wavelengths, which are not observed in our spectra.

\subsection{Measurements of metallicities and ages}
\label{sec:measurement}

We iteratively fit each spectrum to FSPS models via $\chi^2$-minimization. First, we
created a mask for continuum normalization where all emission lines, strong
absorption lines, and telluric regions are masked out. We
continuum-normalized each galaxy spectrum with B-spline fitting
with breakpoints at every $\sim 100$ \angstrom\@. We then created
a different mask for the fitting procedure. In this mask, the Mg~b
triplet and emission lines (if [\ion{O}{2}] was detected in emission) were
masked out from the continuum-normalized spectra. The Mg~b triplet was
masked out so that the measured [Fe/H] would better reflect iron abundance 
rather than magnesium abundance. For each iteration, we simultaneously
fit for four parameters: [Fe/H], age, velocity dispersion, and
redshift. The priors were uniform for [Fe/H] and age. [Fe/H] was in the range of
$[-1.98,0.2]$, while the age was in the range of 0.3 Myr to the age
of the universe at the galaxy's observed redshift. We set the prior
for velocity dispersion according to the Faber-Jackson relation \citep{FaberJackson76} 
between the velocity dispersion and the stellar mass measured by
\citet{Dutton11}, with a range of $\pm0.4$ dex, which is large enough
to capture any uncertainties and evolution with redshift
\citep{Dutton11b}.

The spectral fitting proceeded as follows. First, we used the IDL
code MPFIT \citep{Markwardt12} to fit the continuum-normalized 
observed spectrum with the SSP spectra. In this first fitting iteration, 
all model SSP spectra were continuum-normalized in the same 
manner as the observed spectrum using the
same continuum mask. By fitting continuum-normalized spectra in the
first iteration, we can bypass uncertainties from flux calibration,
dust absorption, and uncertainties of the continuum flux in the
modeled SSPs. This is also important because the MILES library 
provides spectra that have been normalized to unity.

\begin{figure*}
	\centering
	\includegraphics[width=0.9\textwidth]{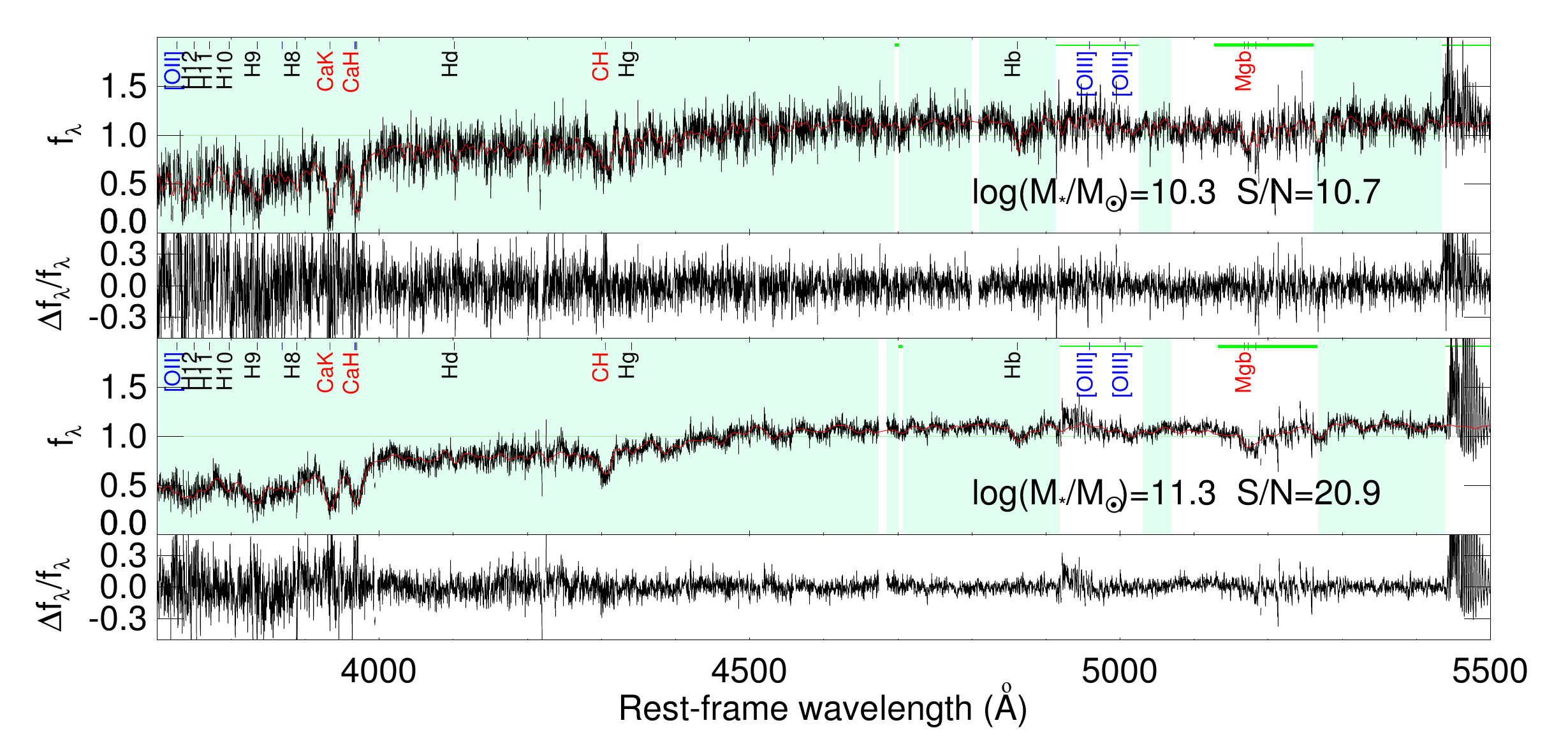}
	\caption{Examples of observed $z\sim0.4$ spectra (black) and
the corresponding best-fit models (red). The flux is
continuum-normalized with an applied ``synthesized'' continuum. The
model spectra are normalized by their median flux and smoothed 
to the instrumental resolution and best-fit velocity dispersion. The teal 
background shows the spectral regions used for spectrum modeling 
while the white background show the spectral regions that are masked 
out. The green bars show regions with strong telluric absorption 
lines. We display the fractional residuals in the bottom
panel of each spectrum. The measured metallicities and ages of the two galaxies are
[Fe/H]$\ =-0.14^{+0.13}_{-0.09}, -0.15^{+0.04}_{-0.06}$ and Age$\ =3.2^{+1.1}_{-0.6}, 6.3^{+0.7}_{-0.8}$ Gyr, respectively.
The uncertainties include the systematic uncertainties from the age--metallicity degeneracy.}
	\label{fig:spec}
\end{figure*}

For the rest of the iterations, we did not continuum-normalize any
model spectra in order to minimize any alteration to the model
spectra. Instead, we applied a ``synthesized" continuum curve to the
observed continuum-normalized spectrum. To do so, (1) we divided the
observed, continuum-normalized spectrum by the best-fit SSP
model spectrum from the previous iteration; (2) we fit the
result from (1) with a B-spline using the same continuum mask; (3) we divided the observed
continuum-normalized spectrum by the continuum curve from
(2) to create an observed spectrum with a ``synthesized" continuum; (4) we
re-fit the resulting spectrum with the SSP model spectra---including the continuum
shape---using the MPFIT code. The process
(1)-(4) was repeated for one hundred iterations, more than enough for
the parameters to converge. We show examples of the observed spectra
and their best-fit spectra in Figure \ref{fig:spec}.

\subsection{Accuracy of metallicities and ages}
\label{sec:accuracy}
In this section, we first examine the statistical uncertainties as a function of
signal-to-noise ratio of the spectra and when the assumption of SSP is dropped.
We then explore the systematic uncertainties that arise from the age--metallicity degeneracy. 
Lastly, we refer the reader to Appendix \ref{sec:compare_measurement} for comparisons 
between our age and metallicity measurements 
and those in the literature. 
\subsubsection{Dependence on signal-to-noise ratios}
\label{sec:sn_dependence}

\begin{figure}
	\includegraphics[width=0.45\textwidth]{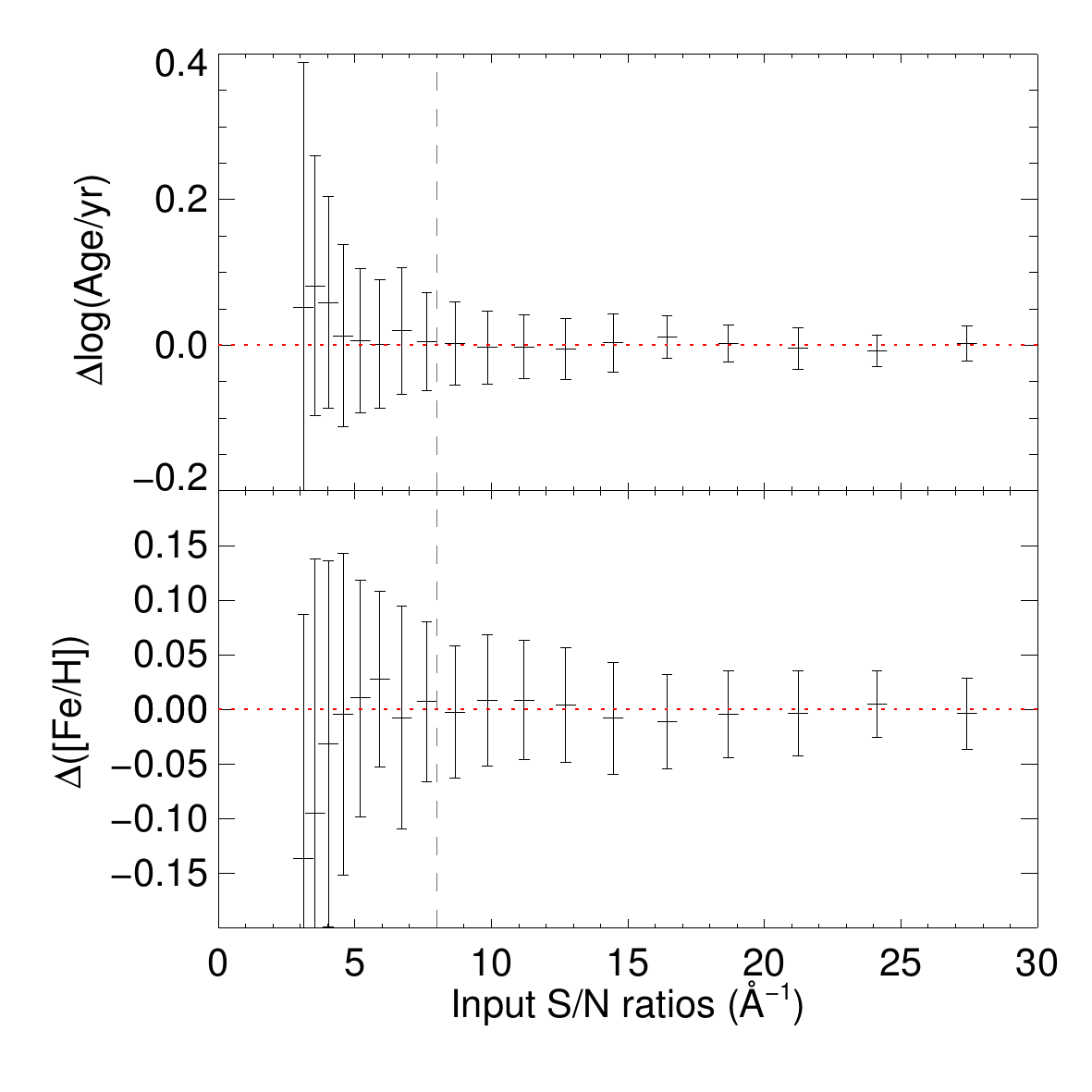}
	\caption{Accuracy in measurements of ages and metallicities as
a function of S/N when assuming a SSP\@. Each data point shows the mean
and standard deviation of 20 mock SSP spectra with a certain S/N\@. At
S/N $\gtrsim 8$ \angstrom$^{-1}$, we can recover the metallicities
within 0.05 dex.}
	\label{fig:SN}
\end{figure}

To investigate how the observed signal-to-noise ratios (S/Ns)
influence the uncertainties in the measured ages and metallicities, we
tested our spectral fitting code on a set of mock DEIMOS spectra with
different S/Ns. We first adopted the SSP assumption. We created a mock
SSP spectrum from the FSPS code with $\log Z=-0.1$ dex and age of 3
Gyr. These numbers were chosen to be representative for our $z\sim0.4$
data. The spectrum was smoothed to have a velocity dispersion of
FWHM $=250$ km/s and the same spectral range and resolution as a
typical DEIMOS spectrum. Gaussian noise was added to the spectrum to
create 20 spectra for each S/N ranging from $\sim 3$ to 30
\angstrom$^{-1}$. We then multiplied the spectra with a telluric
transmission curve and the DEIMOS instrumental throughput to mimic the
observed spectra.

We found that given the SSP assumption, we can measure [Fe/H] and age
well to $\sim0.05$ dex precision when the S/Ns of the spectra are $>8$
\angstrom$^{-1}$, which is the minimum S/N in our sample. 
The uncertainties in [Fe/H] and age as a function
of S/N are shown in Figure \ref{fig:SN}. The fluctuation in the
measurements decreases rapidly as a function of S/N\@. However, once the
S/N is higher than $\sim8$ \angstrom$^{-1}$, the fluctuations
decrease slowly. The level of the fluctuation does not change much
when we change the age of the mock spectra from 3~Gyr to 8~Gyr.
With uncertainties $\lesssim0.05$ dex,
we conclude that we only minimally suffer from statistical
uncertainties.

\subsubsection{Validity of the SSP assumption}
\label{sec:valid_ssp}
Next, we test the validity of the SSP assumption. We created a set of
mock spectra of a composite stellar population (CSP) with an
exponentially declining star formation history $\psi(t')\propto
exp(-t'/\tau)$ with the following parameters: $\tau=1$ Gyr,
$\log (Z/Z_\odot)=-0.2$, and an elapsed time ranging from 1 to 8 Gyr since
the onset of the star formation. We added Gaussian noise so that the
S/N is 12 \angstrom$^{-1}$, the average S/N of our spectra. We applied 
telluric features, the instrumental throughput, and a smoothing kernel 
to the mock spectra in the same manner as in 
Section \ref{sec:sn_dependence}. We note that in these
spectra, all stars have the same metallicity and should be interpreted
as the population's light-weighted metallicity.

\begin{figure}
	\includegraphics[width=0.45\textwidth]{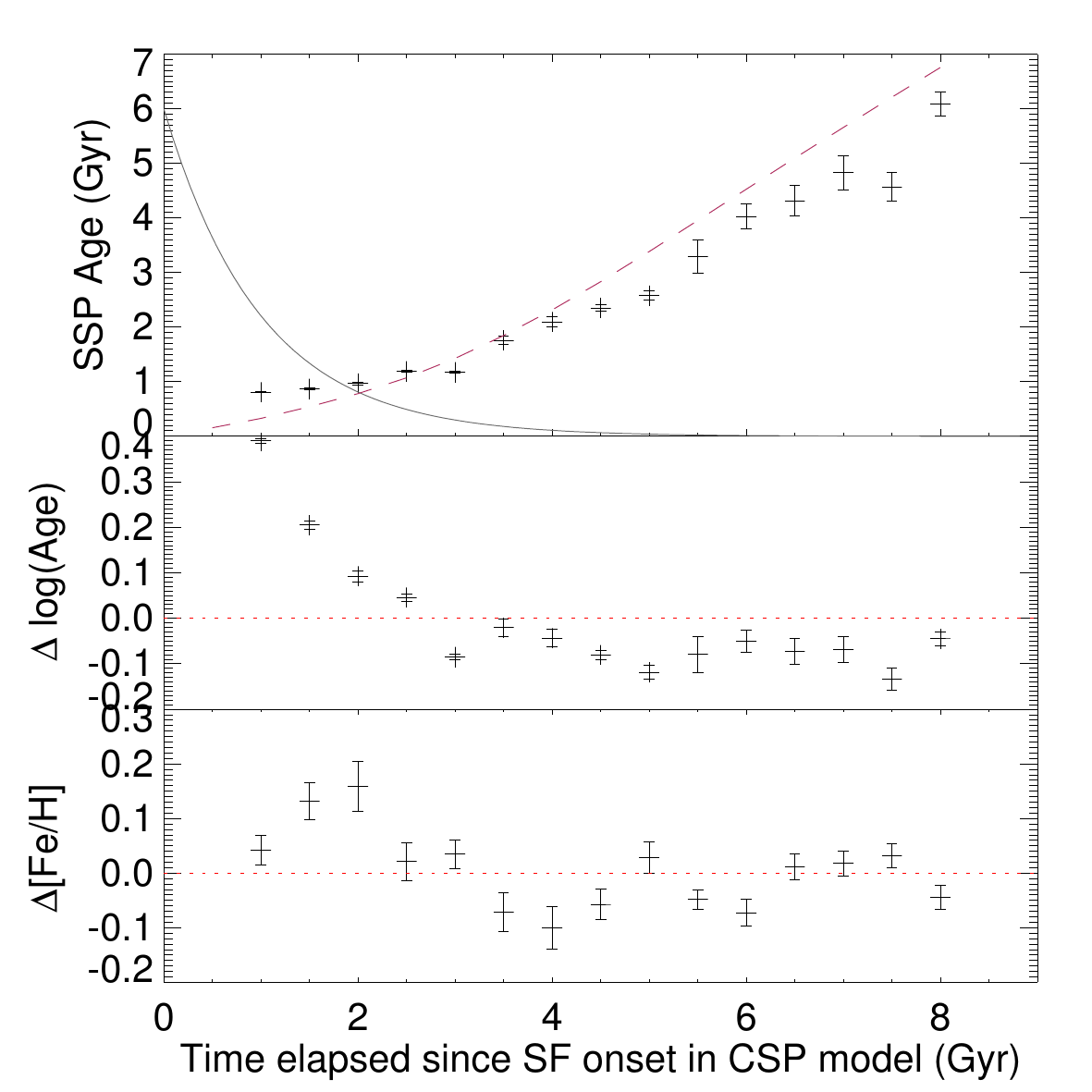}
	\caption{Accuracy in measurements of ages and metallicities
when SSP is not assumed. ({\it Top}) Measured SSP ages from the mock
spectra of composite stellar populations. The spectra have ${\rm S/N} = 12$
\angstrom$^{-1}$ and the same star-formation history but were observed
at different amounts of time elapsed after the onset of the star
formation. The solid curve shows the assumed exponentially declining
star formation history. The red dashed line shows the light-weighted age
calculated over the 3700-5500 \angstrom\ range. 
({\it Middle and bottom}) The deviation of the
measured ages and metallicities from the input values. The middle panel 
shows the difference between the measured SSP ages and the red dashed 
line in the top panel.}
	\label{fig:SN_CSP}
\end{figure}

We then fit the CSP model spectra with SSP models. In general, 
we recovered the CSP age and metallicity within $\sim0.1$ dex precision 
when most of the star formation has been quenched. The measured 
metallicities and ages are shown in Figure \ref{fig:SN_CSP}. The gray 
curve in the top figure shows the shape of the exponentially declining 
star formation rate. The red dashed line is plotted as a guide for the 
``light-weighted'' age of the population at elapsed time $t$ via 
\begin{align*}
\textrm{Age}(t)\sim \frac{\int_{0}^{t}(t-t')L(t')\psi(t')dt'}{\int_{0}^{t}L(t')\psi(t')dt'}   
\end{align*} 
where $L(t')$ is the integrated light in the wavelength range of rest-frame
3700 to 5500 \angstrom\ produced by SSP stars of age $t'$.

Although
the integrated light $L(t')$ accounts for the massive stars that died
before reaching age $t'$, the integration is over every star that has
formed ($\psi(t')dt'$) and does not account for the stars that have
died. Therefore we expect this ``light-weighted'' age to
overestimate the time elapsed since the beginning of star formation, 
especially for older populations. The age overestimate in old populations 
might partially explain the behavior shown in the middle plot. The 
SSP-equivalent ages that we measured are consistently younger than their
respective ``light-weighted'' ages of the CSP population. 

The measured metallicities fluctuate well
within 0.1 dex of the true answer, and they do not seem to 
be affected by assuming an SSP rather than a CSP\@.
We conclude that we can measure metallicity to a precision of
$\sim0.1$ dex under the SSP assumption. On the other hand,
we likely underestimated ages by an amount less than 0.1
dex as long as the majority of the star formation has been quenched,
i.e., the light-weighted age is greater than $\sim1.5$ Gyr. Because our
sample is comprised of quiescent galaxies, we expect that our age and metallicity
measurements are not greatly affected by the SSP assumption.

Our results regarding the SSP assumption are similar to the
results by \citet{TragerSomerville09}. The authors concluded that an
SSP-equivalent metallicity is an excellent tracer of the light- or
mass-weighted metallicities, whereas an SSP-equivalent age generally
biases toward values younger than the true mass- and light-weighted
ages. The level of bias in ages at the level of $\sim-0.1$ dex compared
to the true light-weighted age is also similar to ours. \citeauthor{TragerSomerville09}
argued that hot young stars contribute minimally to the metal lines but
heavily to the Balmer lines relative to old
stars.

\subsubsection{Age--metallicity degeneracy}

\begin{figure*}
	\centering
	\includegraphics[width=0.4\textwidth]{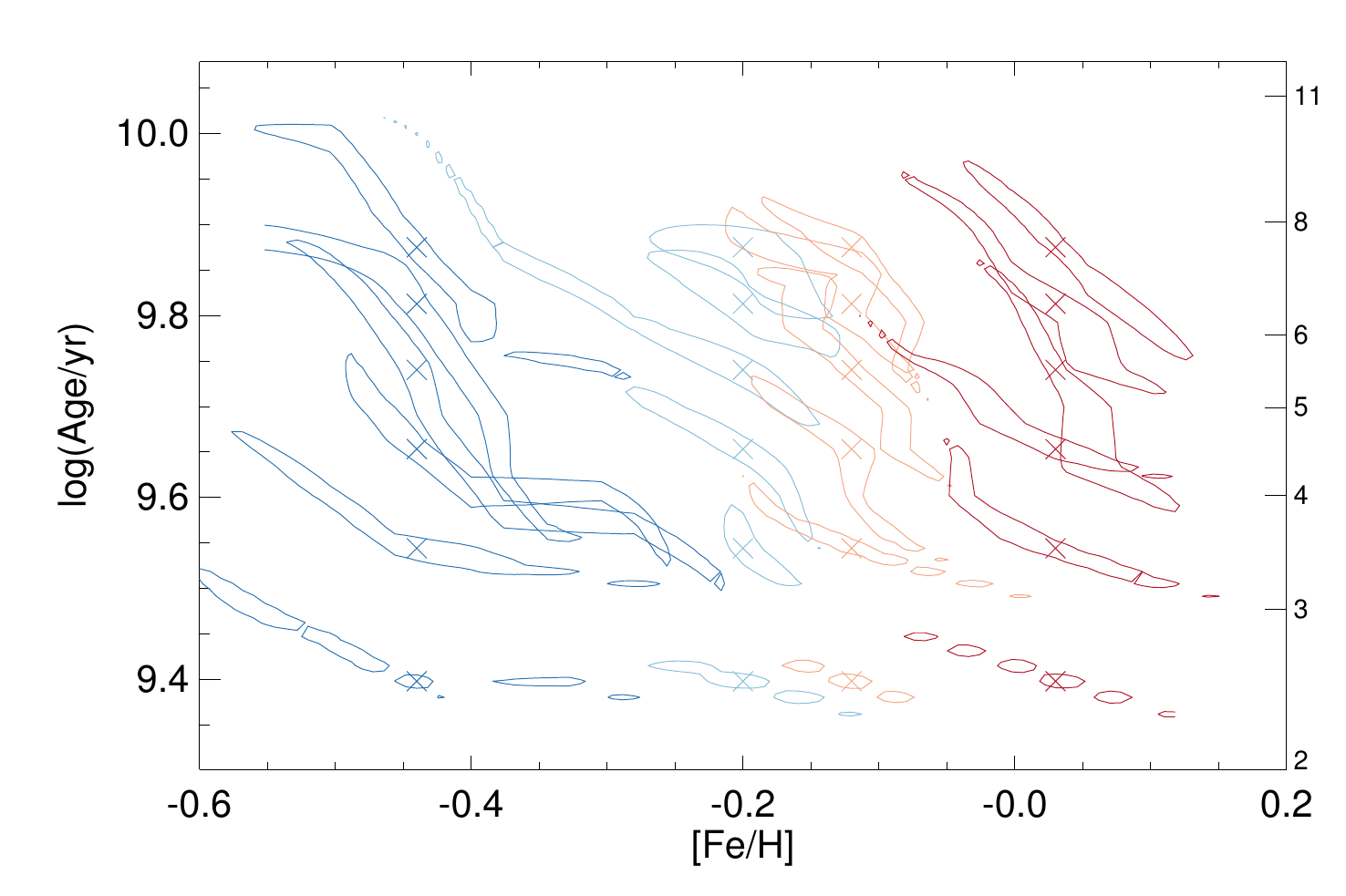}
	\includegraphics[width=0.4\textwidth]{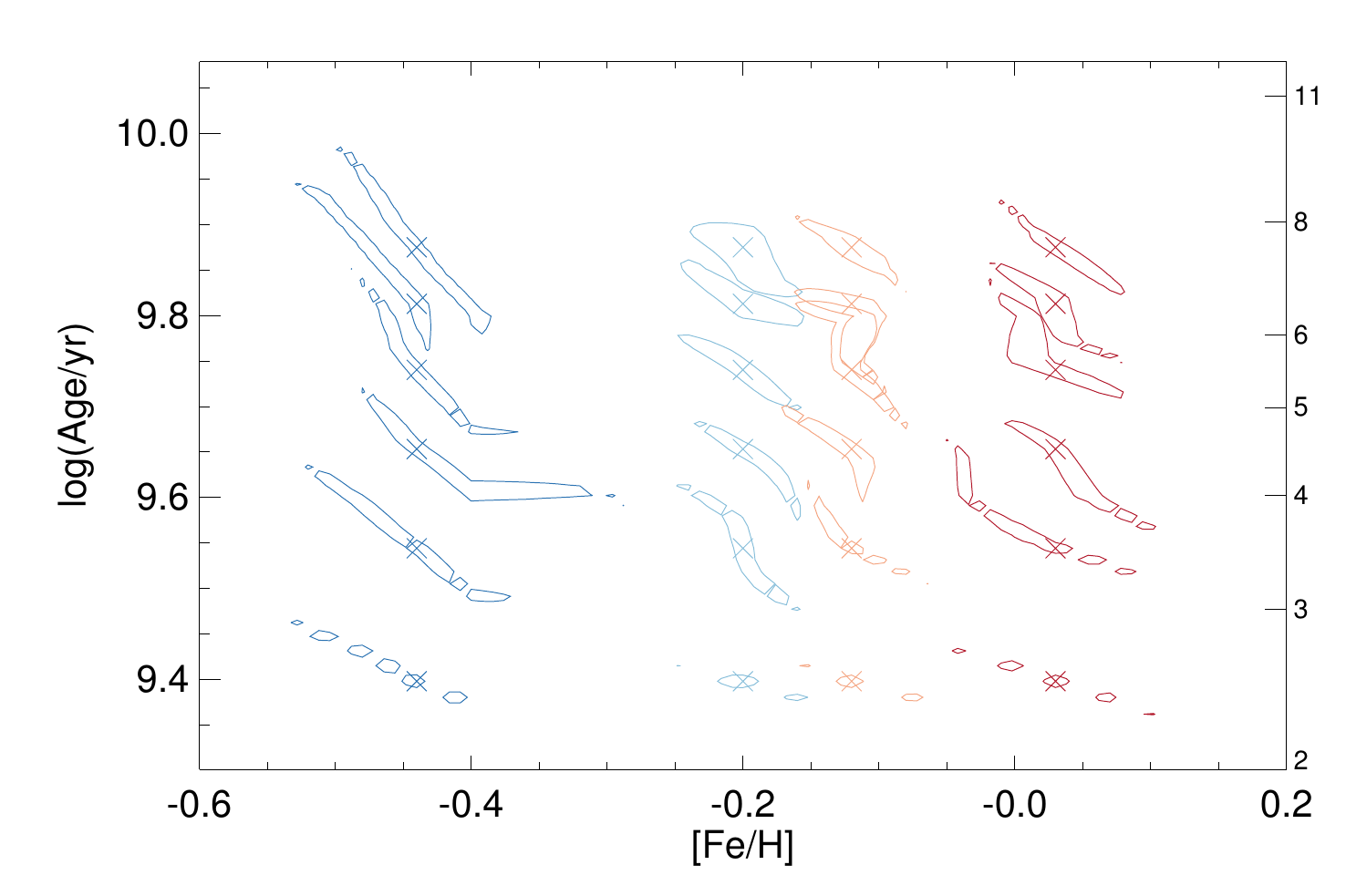}
	\caption{Uncertainties in the measured ages and metallicites
according to the age--metallicity degeneracy. Each contour shows the
1$\sigma$ range in age and metallicity. The true ages
and metallicities are shown as cross marks. The S/N of each spectrum is
10 \angstrom$^{-1}$ (left) and 25 \angstrom$^{-1}$ (right).}
	\label{fig:wormplot}
\end{figure*}

For each galaxy, we estimated the systematic uncertainty arising from 
the age--metallicity degeneracy. As shown in Figure \ref{fig:SN_CSP},
the statistical uncertainties obtained from MPFIT underestimate
the level of total uncertainty. To estimate total uncertainty, we created 
a mock SSP spectrum with the same age and metallicity as each 
observed galaxy. Gaussian noise was added to the spectrum to reach 
the same S/N as the observed spectrum. We then compared the noised 
spectrum to a $100\times120$ grid of noise-less SSP spectra with a 
range [0.5,13] Gyr in age and [-0.8,0.2] dex in [Fe/H] and calculated 
a $\chi^2$ array for the noised mock spectrum. All SSP spectra were 
smoothed to achieve a velocity dispersion of $250$ km/s FWHM 
convolving with the typical resolution of an SDSS or a DEIMOS spectra. 

The uncertainties in age and metallicity are calculated by marginalizing 
the 2--D posterior probability distribution obtained from the $\chi^2$ array.
The uncertainties for upper (lower) limits are the differences between 
the values at 84th (16th) percentile and the 50th percentile in the 
posterior probability distributions. Since we calculate these uncertainties 
specifically to the S/N of each galaxy, we take these uncertainties as 
the total uncertainty of each measurement. The uncertainties quoted in 
the subsequent text and figures refer to these systematic uncertainties. 

In general, the uncertainties in ages and metallicities are neither 
Gaussian nor symmetric. Figure \ref{fig:wormplot} shows contours 
of 1$\sigma$ uncertainty based on the 2--D posterior probability 
distribution functions of the age--metallicity degeneracy at S/N$=10$ 
and S/N$=25$ \angstrom$^{-1}$. The average total uncertainties 
for upper and lower limits are $+0.11$ and $-0.14$ dex for [Fe/H] 
and $+0.12$ and $-0.11$ dex for age, respectively. We list all our 
measurements of ages and metallicities in Table \ref{tab:catalog}.

%%%%%%%%%%%%%%%%%%%%%%%%%%%%%%%%%%%%%%%%%%%%%%%%%%%%%%%%%%%%%
%%%%%%%% SECTION 4 Local MZR %%%%%%%%%%%%%%%%%%%%%%%%%%%%%%%%%%%%%%%%%
%%%%%%%%%%%%%%%%%%%%%%%%%%%%%%%%%%%%%%%%%%%%%%%%%%%%%%%%%%%%%
\section{Mass--Metallicity Relation of Local Galaxies}
\label{sec:localmzr}
In this section, we report the MZR of SDSS local galaxies measured with
our method. We will use this MZR as a reference to compare with the
MZR of $z\sim0.4$ galaxies in Section \ref{sec:main_results}.

\begin{figure*}
	\centering
	\includegraphics[width=0.75\textwidth]{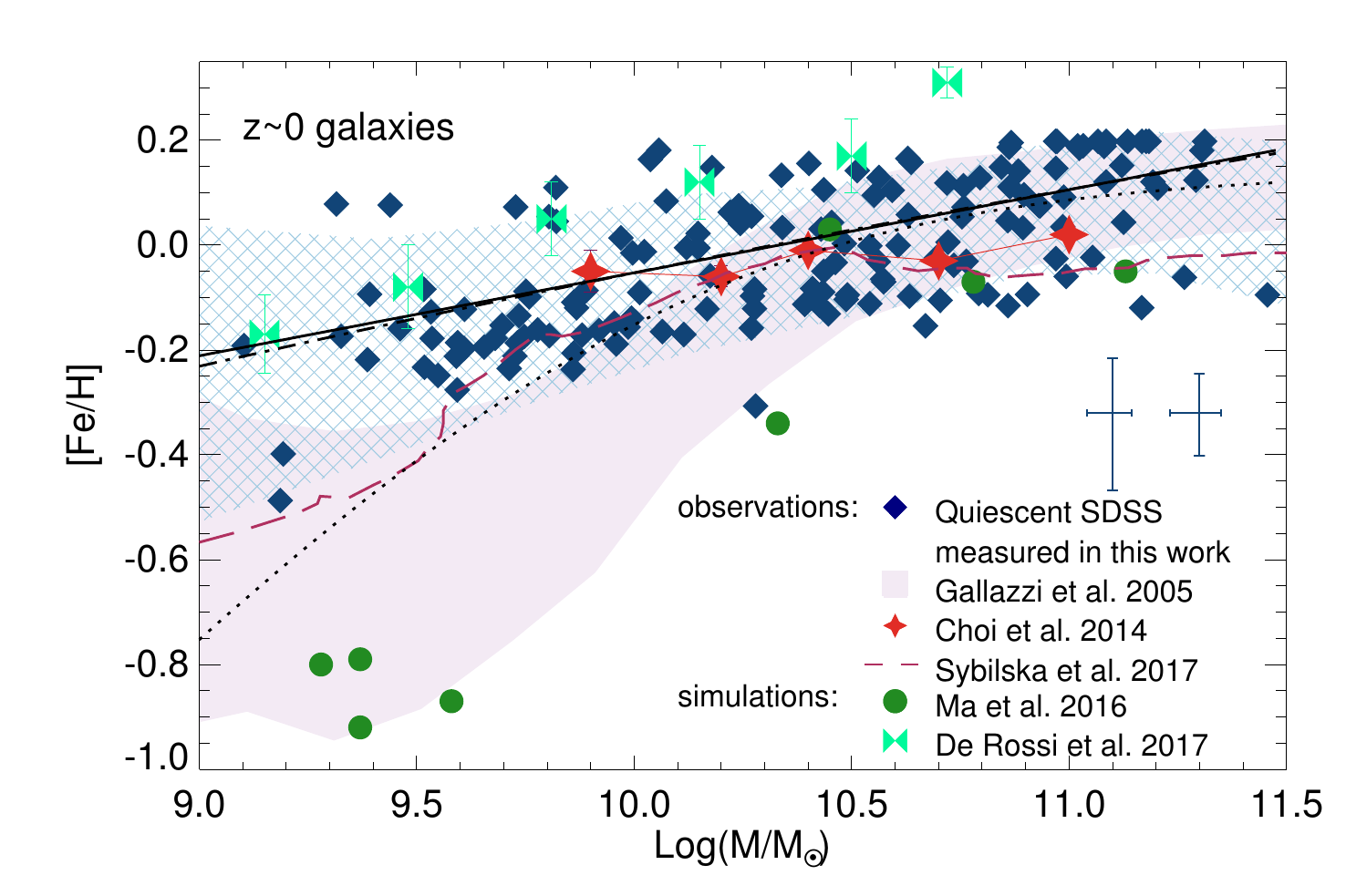}
	\caption{The stellar mass--metallicity relation of local
quiescent galaxies. Each navy-blue diamond shows the measured
metallicities in this work. The light-blue hatched strip shows the
average and standard deviation of metallicities in each mass bin. The
two error bars to the right of the plot show median uncertainties for
galaxies with stellar mass lower (left) and higher (right) than
$10^{10}M_\odot$. The black solid curve is the best-fit linear
function while the black dash-dot and dotted lines and are the best-fit quadratic 
and logarithmic (Equation~\ref{eq:mzrlog}) function, respectively. 
The light-purple solid strip shows the average relation from \citet{Gallazzi05}. 
The plot also shows local MZRs measured in other works \citep{Choi14,Sybilska17} 
and the local MZRs found in the FIRE and EAGLE hydrodynamical simulations
\citep{Ma16,DeRossi17}.}
	\label{fig:MZR_sdss}
\end{figure*}

Figure \ref{fig:MZR_sdss} shows the stellar mass--stellar metallicity
relation we measured for local quiescent SDSS galaxies. We selected 
a subsample of 155 quiescent galaxies from \citet{Gallazzi05}. The 
selection criteria and detailed comparisons to the age and metallicity measurements 
of \citeauthor{Gallazzi05} are described in Appendix 
\ref{sec:compare_measurement}. In order to ensure that 
the stellar masses of both the $z\sim0$ and $z\sim0.4$ samples are on the 
same scale, we remeasured the stellar masses of the 
local galaxies with the KCORRECT code using the SDSS u, g, r, i, z photometry.  
The MZR of local galaxies shows 
the expected relation; metallicity increases with stellar mass. However, 
the ``knee,'' or where the slope in the MZR changes, is less visible 
from our measurements than from \citet{Gallazzi05}. The MZR is 
generally consistent with the MZR of the complete galaxy population
(both star-forming and quiescent galaxies) analyzed by \citeauthor{Gallazzi05} 
at stellar mass above $M_*\approx10^{10.3}M_\odot$. However, at 
lower stellar masses, the MZR of the subsample of quiescent galaxies 
shows metallicities higher than those of \citeauthor{Gallazzi05}.

We consider three functions to approximate the MZR. First, we try fitting with the
three-parameter logarithmic function proposed by \citet{Moustakas11}
to describe the gas-phase MZR:
\begin{equation}
  \text{[Fe/H]}=\text{[Fe/H]}_0-\log[1+M_*/M_0]^\gamma \label{eq:mzrlog}
\end{equation}
We then fit with a quadratic function, another proposed form of gas-phase MZR
\citep[e.g.][]{Tremonti04}, and a linear function. For each function, we fit with 1000 
iterations of the Monte Carlo random sampling method because the uncertainties of our 
measurements are not Gaussian. In each iteration, we resample according to the 
probability distribution in [Fe/H] of each galaxy and find the best-fit function
by minimizing the chi-square error statistic. 

The MZR of the subsample of quiescent galaxies is best 
described by a linear fit with a modest slope. 
The best-fit logarithmic function (the bottom-most dash-dot line in
Figure \ref{fig:MZR_sdss}) performs the worst in terms of $\chi^2$, 
at approximately 1.6 times the minimum $\chi^2$ of the two other functions.
Both quadratic and linear functions fare equally well based on the minimum
$\chi^2$. However, according to the Akaike information criterion
\citep{Burnham03}, the linear function is a better choice because 
it is less complex and minimizes the loss of information. 
The slope of the best linear fit is
$\sim0.16\pm0.02$ dex in [Fe/H] per $\log {\rm mass}$---consistent with the
slope found by \citet{Gallazzi06}. The slope indicates that an increase in galaxy
mass by a factor of 10 corresponds to an increase in metallicity by a factor of 1.4.

When compared with the MZR measured by \citet{Gallazzi05}, 
higher metallicities in the low-mass galaxies can directly result 
from the systematic difference in the metallicity measurements. 
We tried plotting the MZR using the stellar masses reported in 
\citet{Gallazzi05}, which were measured from spectral indices 
and z-band photometry as compared to the KCORRECT code used in this paper. 
Because the mass measurements from both methods are generally consistent within 0.1 dex, 
the best-fit linear function does not depend on the difference in the mass measurement methods. 
The systematic difference in metallicity measurement is therefore the cause of the difference
in the MZR at the lower mass end. 

As discussed in Appendix 
\ref{sec:compare_measurement}, for the same galaxies 
with low metallicities, we measured their [Fe/H] to be higher than 
what \citet{Gallazzi05} measured. Since galaxies with low metallicities 
are mainly less massive galaxies, this results in the MZR lying above 
that shown by \citet{Gallazzi05}. The MZR with higher metallicities 
at the lower-mass end found here is consistent with the MZRs found 
in stacks of local spectra in \citet{Choi14} and in individual local 
early type galaxies via IFU observations by \citet{Sybilska17}. 
The metallicities measured by \citet{Sybilska17} were
measured via spectroscopic indices using the SSP models from
\citet{Vazdekis15}, which is also based on the MILES library. Because
the measurements in \citet{Choi14}, \citet{Sybilska17}, and this work
utilize the same stellar library, it is very likely that the choice of
stellar library used caused the systematic differences in the
metallicity measurements compared to \citet{Gallazzi05}. 

Whether the higher metallicity at the lower-mass end of the MZR can be
additionally caused by differences in SFH between the two samples, 
on top of the measurement methods, is still ambiguous. 
There is observational evidence that
the stellar MZR of early type galaxies might differ from that of
late-type galaxies. Although categorizing galaxies based on their
morphologies is not necessarily the same as categorizing based on
their SFH, the two properties closely correlate
with each other in both local and high-redshift galaxies
\citep[e.g.,][]{Kriek09,Wuyts11,Lee13}. \citet{Gallazzi05} found that
morphology, described by a concentration index, is responsible for a
difference as large as $\sim0.6$ dex in [Fe/H] in low-mass
galaxies. The difference between the two populations in more massive
galaxies is minimal. 

On the other hand, theoretical work suggests that there should not be
a significant difference between star-forming and passive
galaxies. \citet{Okamoto16} created separate MZRs for passive and
star-forming galaxies based on the Illustris simulation
\citep[][Illustris-1]{Nelson15} and the EAGLE simulation
\citep{McAlpine16}. \citet{Okamoto16} showed that the difference between
passive and star-forming galaxies is $\sim0.05$ dex or smaller at any fixed
mass. Furthermore, they found that the shape of the MZR does not depend on galaxy type.

Regardless of the true shape of the MZR at smaller masses, we will use
our measurements in Figure \ref{fig:MZR_sdss} as a point of reference
when comparing with those of higher redshift. Comparing to our own 
measurement of the local galaxy populations reduces the 
systematic uncertainties introduced from measurements and sample selection.

%%%%%%%%%%%%%%%%%%%%%%%%%%%%%%%%%%%%%%%%%%%%%%%%%%%%%%%%%%%%%%%%%
%%%%%%%% SECTION 5 z=0.4 MZR %%%%%%%%%%%%%%%%%%%%%%%%%%%%%%%%%%%%%%%%%%%%%
%%%%%%%%%%%%%%%%%%%%%%%%%%%%%%%%%%%%%%%%%%%%%%%%%%%%%%%%%%%%%%%%%
\section{Evolution in the mass--metallicity relation}
\label{sec:main_results}
In this section we show the main result of this work. In Section
\ref{sec:cl_results}, we report the MZR
of individual $z\sim0.4$ galaxies in which we, for the first time, 
detect an evolution in the stellar MZR with
observed redshift. In Section \ref{sec:evolution_formz}, we explain the scatter
in the MZR and report an even greater evolution of the MZR when the age of
galaxies is taken into account. Lastly, in Section
\ref{sec:slope}, we discuss the meaning of the MZR slope and what it
reveals about feedback in galaxies.

\subsection{Evolution of the MZR with observed redshift} 
\label{sec:cl_results}
\begin{figure*}
	\centering
	\includegraphics[width=0.75\textwidth]{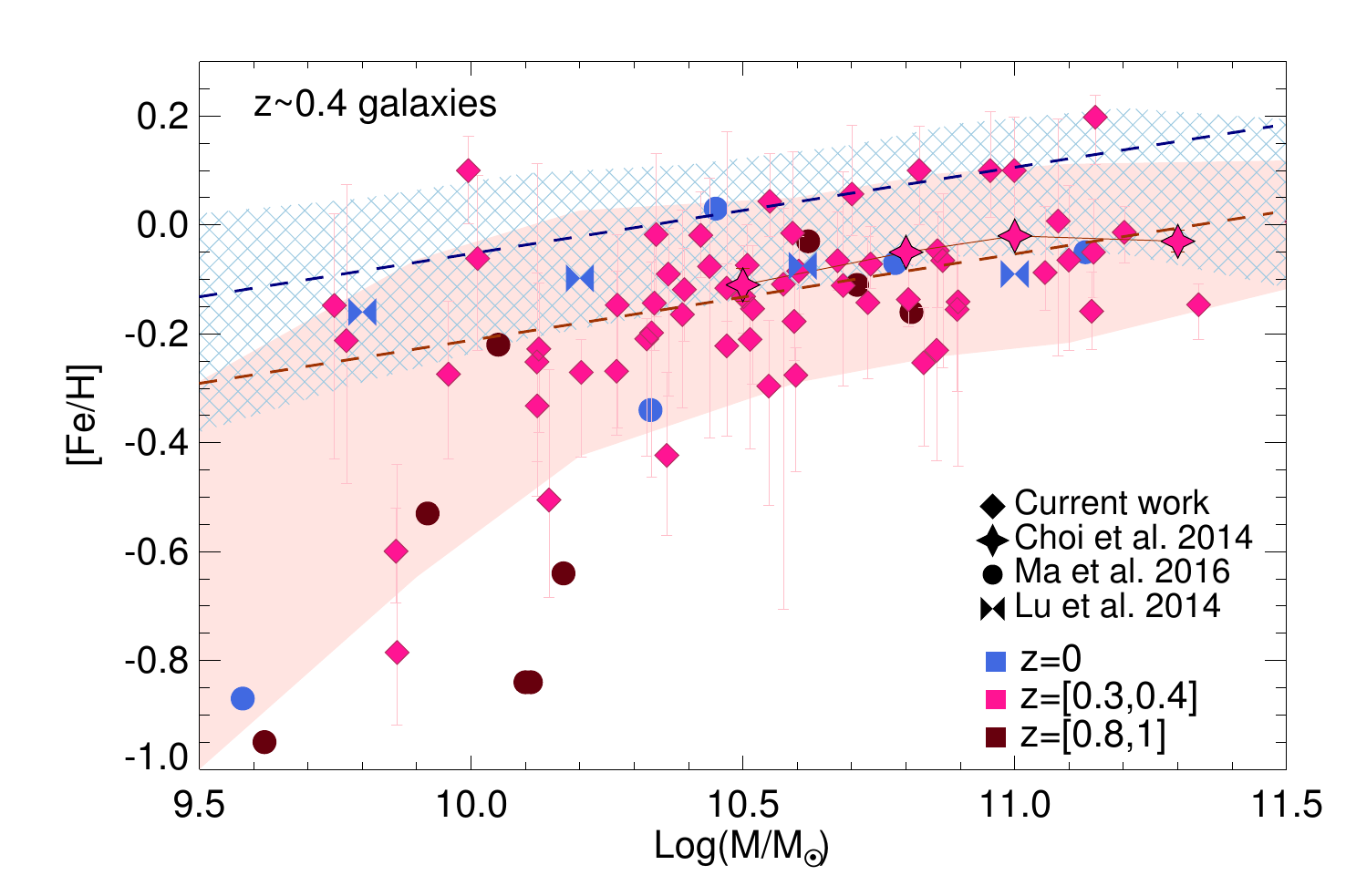}
	\caption{Stellar MZR relation of $z\sim0.4$ quiescent
galaxies. The solid orange and blue hatched strips show the average and
standard deviation of metallicities in each mass bin of $z\sim0.4$ and
local galaxies (same as Figure \ref{fig:MZR_sdss}). The upper dashed
line and lower dashed line show the best-fit linear functions, where slopes 
were fixed to the common value, to the local and $z\sim0.4$ galaxies, 
respectively. The best-fit parameters are shown in Equation \ref{eq:z0} and 
\ref{eq:z04}. We also show the predicted MZR from 
the FIRE simulations \citep{Ma16} and from semi-analytic models 
with a constant mass-loading factor \citep{Lu14}. 
The plot is color coded by galaxy redshift.}
	\label{fig:MZR_obs}
\end{figure*}

We plot the stellar MZR of the $z\sim0.4$ sample in Figure \ref{fig:MZR_obs}. 
The $z\sim0.4$ MZR matches well with that measured by \citet{Choi14} 
at the same redshift at $M_*>10^{10.5}M_\odot$. However, in contrast to 
\citet{Choi14}, we measured the MZR of individual (rather than stacked) 
galaxies and extended the MZR relation almost 1 dex lower in stellar mass, 
at $10^{9.7}M_\odot$.

Based on the best-fit linear relations to the MZRs at $z\sim0$ and 
$z\sim0.4$, we find an evolution of the MZR with observed redshifts 
at greater than a $5\sigma$ significance level. As in Section \ref{sec:localmzr}, 
we fit a linear function to the measured $z\sim0.4$ MZR via the Monte 
Carlo method. We then use the analysis of covariance to compare 
the best-fit linear functions of the MZR at $z\sim0$ and $z\sim0.4$. 
First, we check if the slopes of the the two linear functions are different. 
The best-fit slope for the $z\sim0.4$ population is 
$0.15\pm0.03$ dex per $\log {\rm mass}$.
When compared to the slope of $0.16\pm0.02$ dex per $\log {\rm mass}$ 
of the $z\sim0$ population, they are the same within $\sim1\sigma$ significant. 
Therefore, we conclude that the slopes of the MZRs at two redshifts 
are not significantly different. We then test for the evolution in the 
normalization values (the constant terms). To do so, we re-fit linear 
functions to the two MZRs using a common fixed slope, equal to 
the weighted-mean slope of $0.16$ dex per $\log {\rm mass}$. 
The best-fit linear equations when the slopes are fixed are 
\begin{equation}
\langle\textrm{[Fe/H]}\rangle = (-0.05\pm0.01)+0.16\log\Big(\frac{M_*}{10^{10}M_\odot}\Big)
\label{eq:z0}
\end{equation}
for the MZR of $z\sim0$ quiescent galaxies, and 
\begin{equation}
\langle\textrm{[Fe/H]}\rangle = (-0.21\pm0.02)+0.16\log\Big(\frac{M_*}{10^{10}M_\odot}\Big)
\label{eq:z04}
\end{equation}
for the MZR of $z\sim0.4$ quiescent galaxies. The difference in the constant terms is
$0.16\pm0.03$ dex, which is significant at greater than a $5\sigma$ level. 

The observed metallicity evolution is larger than but consistent with 
the predictions from hydrodynamical simulations. For now, we 
focus on the shifts in the MZRs with redshift and ignore the slopes  
predicted in simulations when comparing to our observations. 
The observed metallicity evolution of $0.16\pm0.03$ dex from
$z\sim0.4$ to $z\sim0$ translates to an increase of metallicity at 
$0.037\pm0.007$ dex per Gyr. Based on the FIRE hydrodynamical  
simulations, \citet{Ma16} predicted that the stellar metallicity evolution 
from $z=0.4$ to $z=0$ should be 0.13 dex in [Fe/H]. The observed 
evolution is slightly larger than but consistent within $1\sigma$ with the evolution 
predicted from the FIRE hydrodynamical simulations. We found a 
larger discrepancy when comparing the observed evolution to the predicted
evolution from \citet{DeRossi17} base on the EAGLE hydrodynamical simulation.  
 \citeauthor{DeRossi17} predicted evolution of $0.11\pm0.09$ dex from $z=1$ to 
$z=0$ in $10^{10}M_\odot$ galaxies (see their Figure 5), 
which is $\sim0.014\pm0.012$ dex per Gyr. Our observed 
evolution is therefore also greater but, due to the large uncertainty in 
the predicted evolution, is consistent within $2\sigma$.

The observed evolution of stellar MZR likely emphasizes the 
importance of metal recycling in galaxies. Among the hydrodynamical 
simulations, the FIRE simulations predicted the strongest evolution 
of the stellar MZR over redshift \citep{Ma16}, and is the most 
consistent with our observations. As explained in \citet{Ma16}, 
this is because the simulated galaxies in the FIRE simulations 
are able to retain more metals, resulting in higher increase in 
metallicity over redshifts. This is particularly true in 
galaxies with stellar mass above $10^{10} M_\odot$, where the 
retained fractions of metals in the halos are almost unity. 
The reason that different simulations achieve different metal retention 
fractions lies in the physical models on which the simulations 
are based. Many of the cosmological simulations 
(including the EAGLE simulations) adopt `sub-grid' 
empirical models of galactic winds and stellar feedback, where 
fractions of gas are forced to escape the galaxy due to energy 
injection from supernovae and stellar winds. In contrast, the FIRE simulations 
adopt a Lagrangian formulation of smooth particle hydrodynamics 
\citep{Hopkins14}, where metallicities are derived from tracked individual 
star particles that can be ejected and, importantly, accreted back to the galaxy. 
As a result, the metal recycling effect is better captured in the FIRE simulations 
than those that assume `sub-grid' models.

However, complex and more realistic simulations such as the FIRE simulations are 
computationally expensive and still limited in terms of sample size, which 
might explain the small discrepancy between the observed and predicted 
amount of evolution. The predicted evolution of 0.13 dex from $z=0.4$ to $z=0$
in \citet{Ma16} came from fitting a linear function to the MZRs over 
a wide mass range, from $M_*\sim10^{4}M_\odot$ to $M_*\sim10^{11.5}M_\odot$ 
at different redshifts. If we limit the mass range to above $10^{9.7}M_\odot$, 
as in our observations, there are only 4 simulated galaxies at $z=0$ 
and 8 simulated galaxies at $z=0.8$. With this limited mass range and number 
of sample size, the evolution of the MZR in the simulated galaxies is  
is $0.2\pm0.6$ dex at $10^{10}M_\odot$ from $z=0.8$ to $z=0$ 
or $0.03\pm0.09$ dex per Gyr. 
Moreover, the FIRE simulations do not include feedback from possible 
active galactic nuclei (AGNs), which can potentially affect the metallicities 
in massive galaxies with $M_*>10^{11}M_\odot$ \citep{Ma16}. 
Although the evolution from 
$z=0.8$ to $z=0$ in the FIRE simulation is not significant, it is 
consistent with the better constrained values from our observations.

We note that the galaxies in both the EAGLE and the FIRE simulations 
are not necessarily passive, whereas galaxies in our sample are. The fact that the 
observed evolution of the MZR with redshift is consistent with the simulations 
does not have any implication on whether the two populations' metallicities are the same at 
any given redshift (the latter has been suggested by \citet{Okamoto16}). In this 
section, we only compared the magnitude of the change of metallicity 
with redshift, but not the metallicities themselves.
In fact, none of the metallicity values are consistent. 
Different suites of simulations predict different MZR normalizations 
at each redshift, none of which are consistent with each other or with 
our observations (see Figures \ref{fig:MZR_sdss} and \ref{fig:MZR_obs}). 

\begin{figure*}
	\centering
	\includegraphics[width=0.49\textwidth]{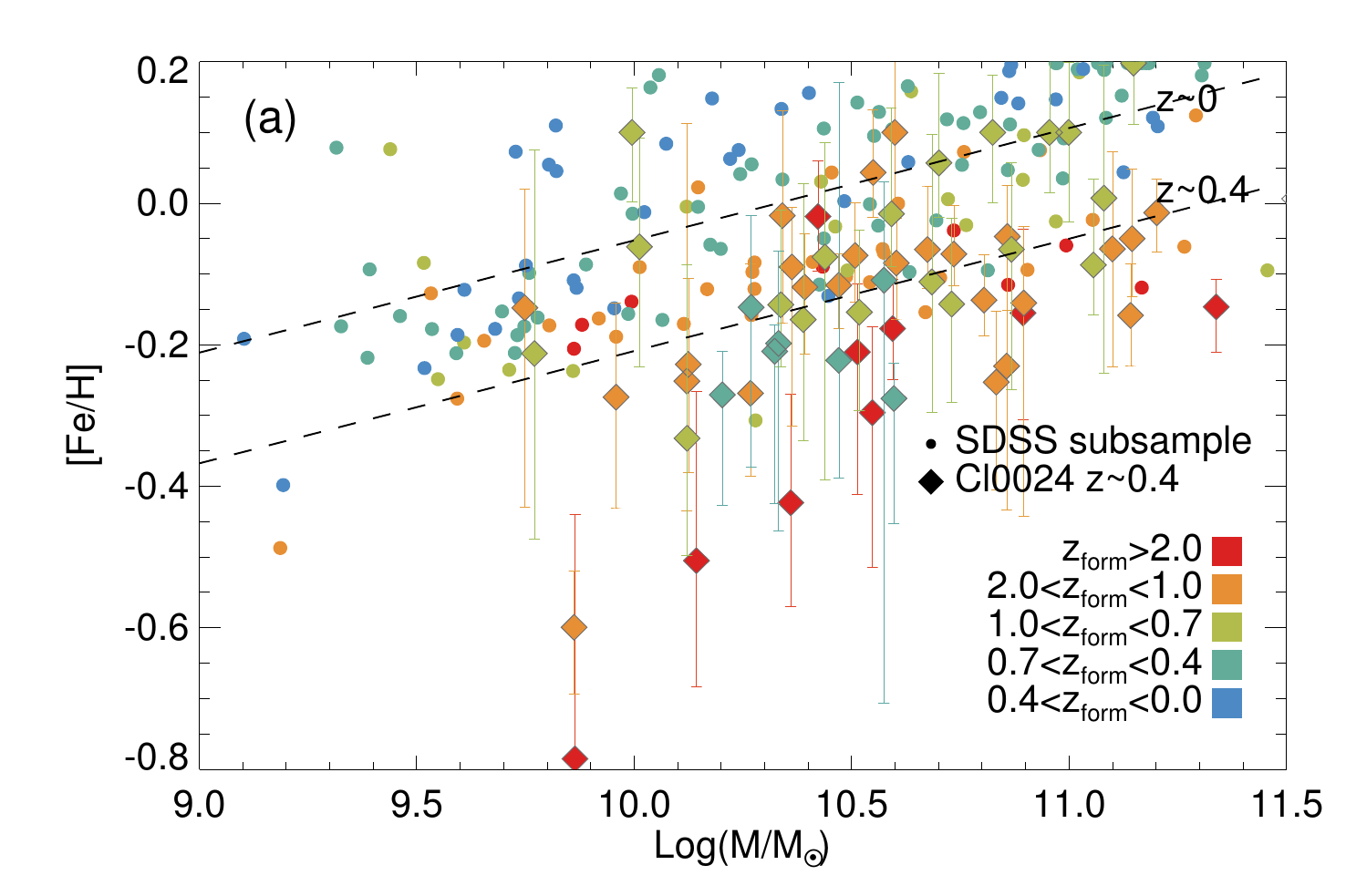}
	\includegraphics[width=0.49\textwidth]{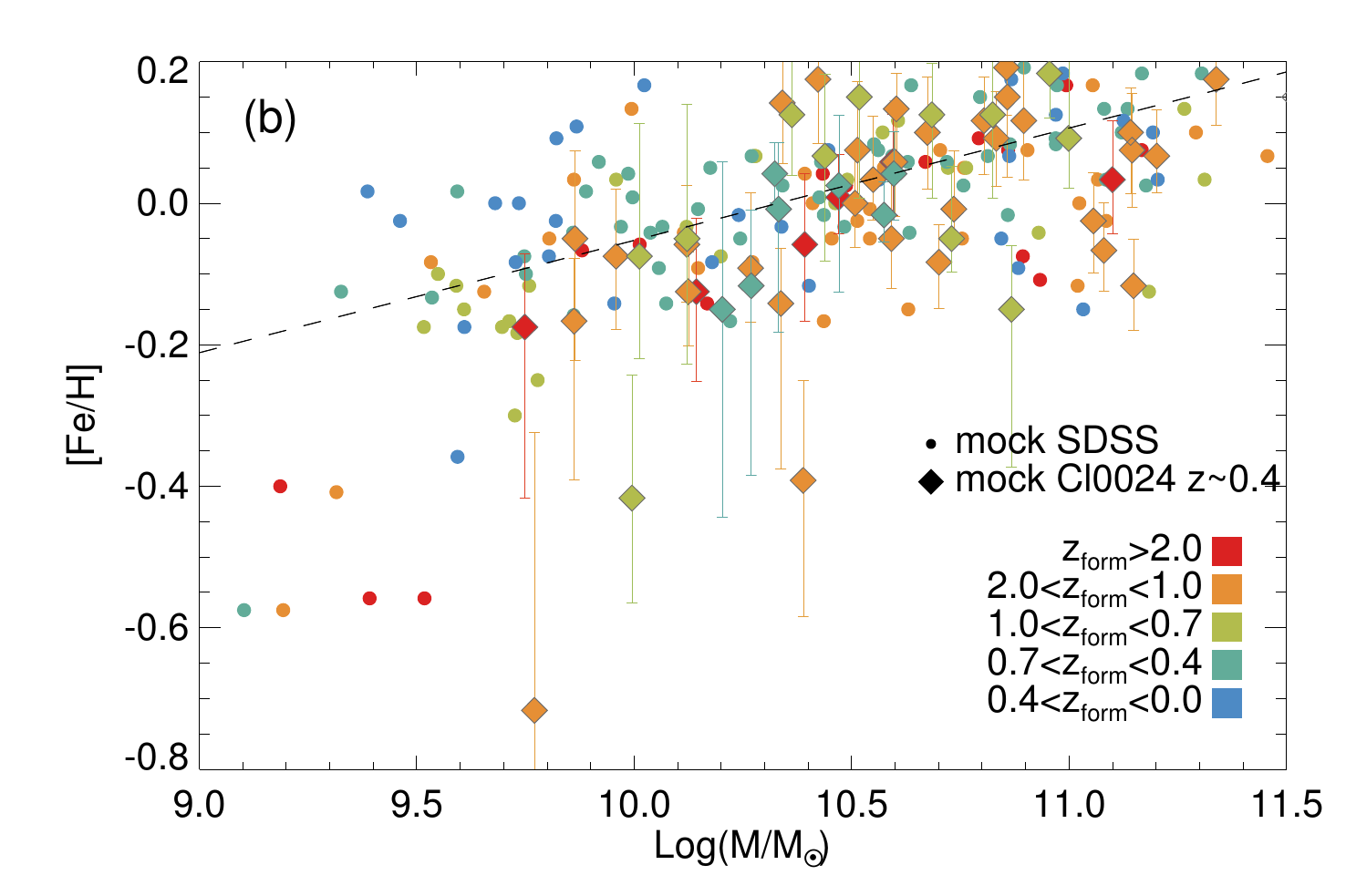}\\
	\includegraphics[width=0.49\textwidth]{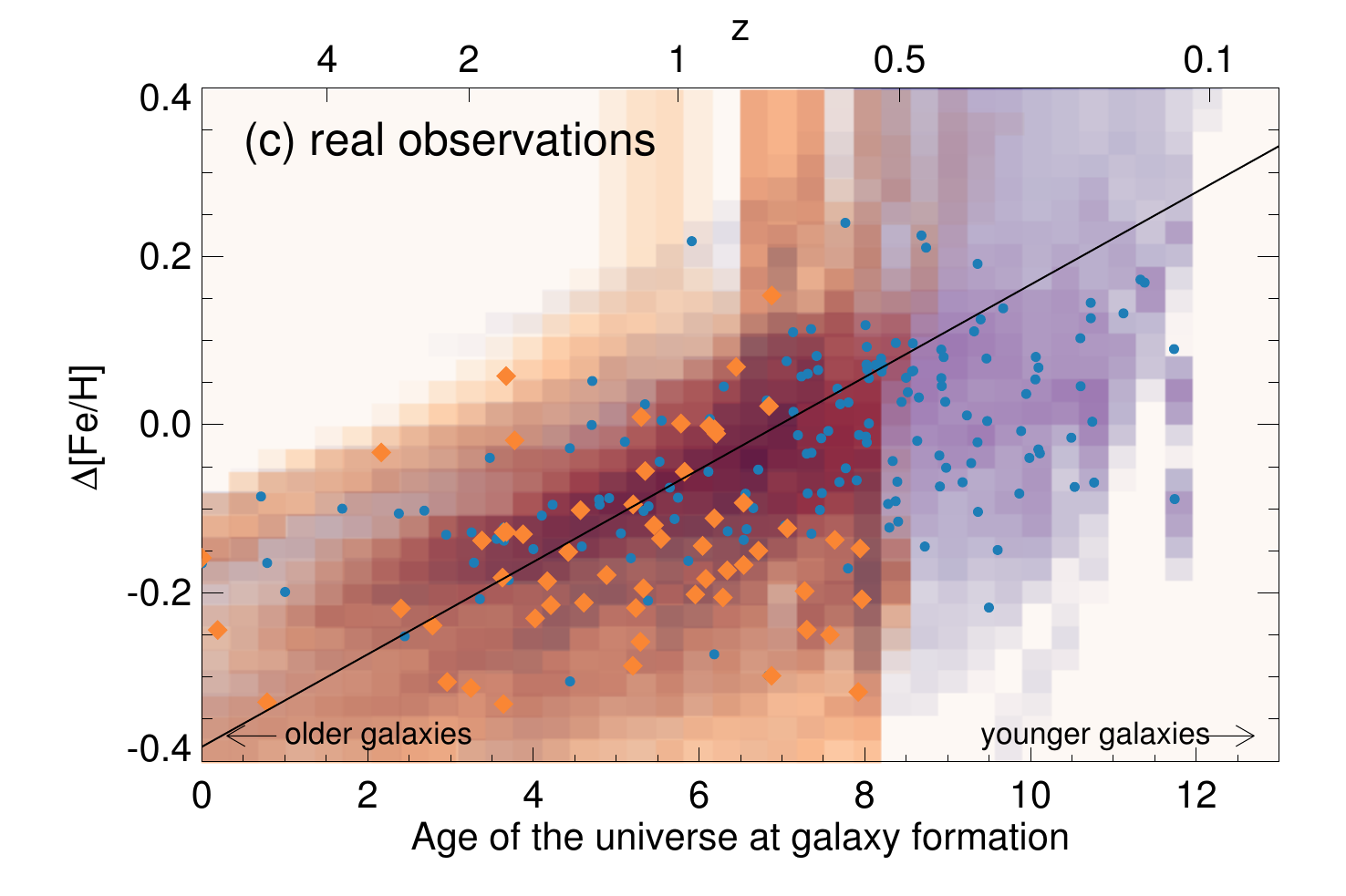}
	\includegraphics[width=0.49\textwidth]{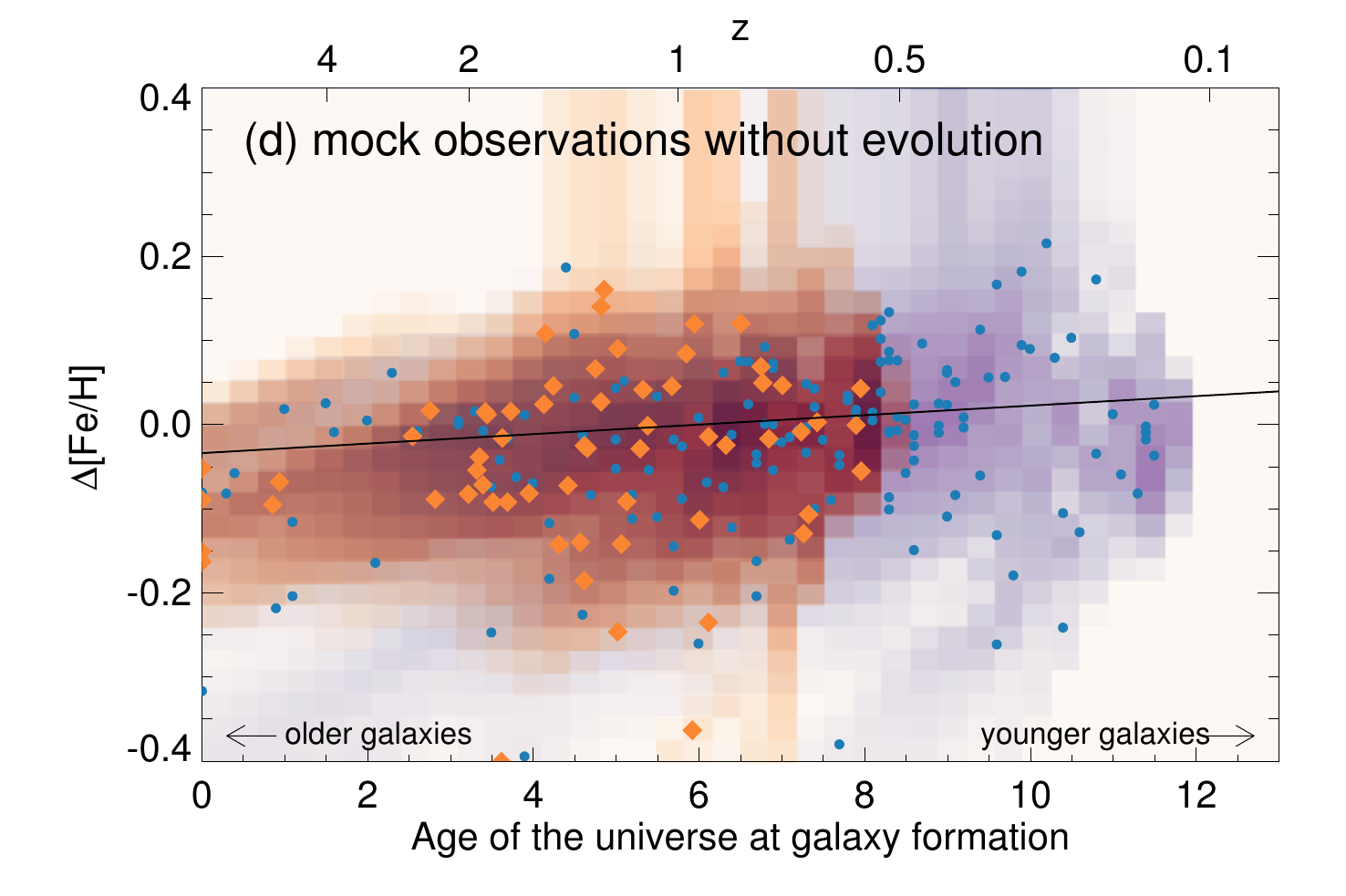}
	\caption{(a) Stellar MZRs of both local and $z\sim0.4$
galaxies color-coded by their SSP-equivalent formation redshifts. 
The black dashed lines show the best-fit linear functions 
of the data at $z\sim0$ and $z\sim0.4$. (b) Mock observation when the ``true" metallicity of each data
point is assumed to be a linear function of its mass (the $z\sim0$ black dashed
line). The ``observed'' ages and metallicities of the mock galaxies were
determined by the age--metallicity degeneracy. (c) and (d) The
deviation of metallicities from the $z\sim0$ best-fit linear function in
observed data and mock data. The orange diamonds and the blue dots represent 
data points from $z\sim0.4$ and $z\sim0$, respectively. The underlying shaded 
purple (orange) colors are the co-added probability distribution of individual data points 
from $z\sim0$ $(z\sim0.4)$ sample. The slope in the mock data (d) is caused
by the age--metallicity degeneracy and is significantly smaller than
the slope in the observation (c), which suggests an evolution of the
MZR with formation redshift.}
	\label{fig:formation_mzr}
\end{figure*}

\subsection{Can galaxy formation time explain the evolution of the MZR with observed redshift?}
\label{sec:evolution_formz}
We now measure the intrinsic scatter in the MZR and test for
the correlation between the scatter and the galaxy's redshift of
formation.

We repeat the linear fit to the stellar MZR with
an additional parameter, an intrinsic variability $\sigma_v$, by
minimizing the negative-log-likelihood 
$$L=-\sum_{i}\log\big[P(\Delta\textrm{[Fe/H]}_i)*\mathcal{N}(0,\,\sigma_v^{2})\big]\big|_{\Delta\textrm{[Fe/H]}_i=0}$$
where $P(\Delta\textrm{[Fe/H]}_i)$ is  the probability of the 
difference between the observed metallicity and the model 
linear equation $\textrm{[Fe/H]}_i-(a\log \frac{M_{*,i}}{10^{10}M_\odot}+b)$, 
which has the same shape as the probability of each observed 
metallicity $P(\textrm{[Fe/H]}_i)$. This equation means that the 
probability of the deviation of each observed metallicity from the 
linear model is equal to the convolution between the probability 
function inherited from the age--metallicity degeneracy 
$P\textrm{[Fe/H]}_i$ and the Gaussian probability of an 
intrinsic scatter of size $\sigma_v$. 

The intrinsic scatter in [Fe/H] for local and $z\sim0.4$ 
sample are both 0.07$\pm0.01$ dex.
The sample at both redshifts have consistent intrinsic
scatter within uncertainties. The measured intrinsic scatters 
do not change when we leave the slope and intercept (a and b) 
as free parameters or fix them to the values in Equations \ref{eq:z0} 
and \ref{eq:z04}. Interestingly, this level of scatter is slightly smaller
than the intrinsic scatter of $\sim0.1$ dex found in the gas-phase MZR
\citep[e.g.][]{Yabe12, Guo16} for galaxies with $M_*\gtrsim 
10^{9.5} M_\odot$ but comparable to the intrinsic
metallicity scatter of $\sim0.05-0.08$ dex in the
fundamental metallicity relation or in the gas-phase MZR when the
star formation rate is taken into account
\citep[e.g.][]{Mannucci10,Yates12,Lilly13}. This is expected because the
stellar metallicity is less affected by the current star formation
rate.

To further investigate the source of the intrinsic scatter, we plot
the MZRs derived from both local and $z\sim0.4$ galaxies, 
color-coded by their formation redshifts, in Figure
\ref{fig:formation_mzr}a. Ideally, this should yield the 
MZR of the star-forming galaxies at each formation redshift, 
which should not, or at most weakly, depend on when 
the galaxies were observed, i.e. the observed redshifts. 
Remarkably, the figure shows that, at each fixed mass, 
the galaxies that formed earlier (red data points) generally 
have lower metallicities than galaxies that formed later 
(blue data points) do. This is as we expect from the evolution 
of the gas-phase MZR\@, if stars approximately adopt the 
metallicity from their birth clouds.

At this point, we conjecture that the stellar MZR does not 
only depend on galaxy mass, but also on the redshift that 
galaxy formed (or when the majority of stars formed 
when the SSP is not assumed.) If this is true, the dependence on 
galaxy formation time should be able to explain the observed evolution 
with observed redshift. To see this effect better, 
we plot the deviation of the measured metallicities from the 
best linear fit of $z\sim0$ galaxies (Equation \ref{eq:z0}, the 
upper dashed line) as a function of the age of the universe at 
the formation of their stellar populations in Figure \ref{fig:formation_mzr}c. 
By subtracting off the mass-metallicity function from the observed 
metallicities, the figure shows the effect of the age of universe at galaxy formation on 
stellar metallicity when the dependence on mass is removed. 
We can clearly see that galaxies that formed earlier (older galaxies) 
offset toward lower metallicities, while galaxies that formed later 
(younger galaxies) offset toward higher metallicities.

However, the correlation with formation redshift (older galaxies
have lower metallicities) is in the same direction as the
age--metallicity degeneracy. Though some previous works have pointed
out the anti-correlation between age and metallicity at a given mass,
it was complicated by or was thought to be the result of the
age--metallicity degeneracy \citep[e.g.,][]{Jorgensen99,
Gallazzi05}. To test whether the scatter in the MZR as a function of
formation redshift is real or caused by the age--metallicity
degeneracy, we created a set of mock observed ages and
metallicities. We assume that the metallicities are solely
determined by stellar mass according to the best-fit linear
function found in the observed MZR at $z\sim0$. If the trend with age is caused
by the age--metallicity degeneracy, then we should obtain the same
level of scatter in the MZR and its correlation with formation 
redshift after noise is added.

For each observed galaxy, we construct its twin mock 
galaxy. We took the measured mass of each observed 
galaxy and calculated its ``true'' metallicity from the linear 
function found in Section \ref{sec:cl_results} 
(Equations \ref{eq:z0}). We also took the 
measured age as the ``true'' age of the galaxy. The ``true'' 
observed spectrum was obtained from the FSPS according 
to its ``true'' age and metallicity, smoothed to the observed 
velocity dispersion convolved with SDSS/DEIMOS instrumental 
dispersion. Gaussian noise was added at each pixel 
with the same flux uncertainty array from the observed 
spectrum. We calculated a $\chi^2$ grid 
for each noised spectrum with noiseless SPS spectra 
of every possible age and metallicity combination in the 
grid of 0.5 to 13 Gyr in age and $-0.8$ to $0.2$ dex in 
metallicity. The grid spectra were smoothed to the 
same dispersion as the noised spectrum. The ``observed" 
age and metallicity of each mock spectrum was then 
selected according to the probability of each cell in
the $\chi^2$ grid. The resulting mock MZR is shown in 
Figure \ref{fig:formation_mzr}b.

We do not find the same level of separation of the mock MZR with
galaxy formation redshift as in the observed MZR\@. The dots and
diamonds of different colors in Figure \ref{fig:formation_mzr}b are
visibly more mixed than those in Figure \ref{fig:formation_mzr}a. 
We plot the deviation of the ``measured" metallicities from the best 
linear fit to the $z\sim0$ stellar MZR as a function of the age of 
the universe at their formation (Figure \ref{fig:formation_mzr}d). 
The slope in Figure \ref{fig:formation_mzr}d (mock observation) 
is purely caused by the age--metallicity degeneracy. If there 
were no age--metallicity degeneracy,
the deviation from the best-fit linear relation should scatter
around $\Delta$[Fe/H]=0 at all ages. However, the degeneracy
causes the data points to move sligthly toward the lower left (more metal-poor
and older) or upper right (more metal-rich and younger). Fitting a
linear fit to the underlying probability distribution with a maximum 
likelihood estimation \footnote{Because the uncertainties of individual ages
and metallcities are not Gaussian and highly correlate with each other 
(see Figure \ref{fig:wormplot}), we cannot use linear-fit estimators that 
assume Gaussian probability distribution. In this case, we use 
Markov chain Monte Carlo sampling to obtain the best linear fit that 
minimize the negative likelihood, $-\sum\log(\oint P_idl(a,b))$, where 
$P_i$ is the probability distribution of individual measurements of 
$\Delta$[Fe/H] and age of the universe at galaxy formation. 
The integration is along the considered linear function 
with parameters a and b for the slope and intercept. The summation is over all data points.
The best linear fit is the line that passes through the highest probability regions. Note that 
the underlying distribution in Figure \ref{fig:formation_mzr}c and \ref{fig:formation_mzr}d
are $\sum P_i$, which represent uncertainties of the data.}, we
found a small positive slope of $0.005\pm0.003$ dex per Gyr in the
relation in Figure \ref{fig:formation_mzr}d.

The slope in the relation between the observed deviation 
from the best-fit line in the MZR as a function of formation 
redshift is steeper than that of the mock observation. 
The slope of the deviation in Figure \ref{fig:formation_mzr}c 
is $0.055\pm0.006$ dex per Gyr, significantly larger than 
the slope in Figure \ref{fig:formation_mzr}d. When 
we fit linear functions to the $z\sim0$ and the $z\sim0.4$ sample individually, the 
best-fit slopes and intercepts are consistent within uncertainties. 
(The slopes are $0.058\pm0.010$ and $0.48\pm0.07$ dex per Gyr for the $z\sim0$ 
and $z\sim0.4$ sample, respectively.) This confirms 
that the evolution of the MZR with formation redshifts is
real. Galaxies with SSP-equivalent formation redshift at 
$z\sim2$ have [Fe/H] on average of 0.4 -- 0.5 dex lower than 
the metallicities of galaxies that just formed in the past 2 Gyr.

The evolution of the MZR with formation redshift suggests 
that the mass--metallicity relation is not only determined 
by galaxy masses but also star formation histories. 
Moreover, this evolution of the MZR with formation
redshift can consistently explain the evolution with observed 
redshift found in Section \ref{sec:cl_results}, which was 
$0.16\pm0.03$ dex from $z\sim0.4$ to $z\sim0$. The difference 
in the weighted mean formation time of the two populations is $2.7\pm0.1$ Gyr. 
If this difference is multiplied by the evolution with formation redshift 
$0.055\pm0.006$ dex per formation Gyr, we would expect a 
$0.15\pm0.02$ dex difference in [Fe/H] between the two 
populations, consistent with what we observed in the evolution 
with observed redshift. The gentler evolution of the MZR with 
observed redshift compared to the evolution with formation redshift 
is probably the result of a shared histories prior to quenching 
that smear out the evolution. 

Lastly, we note that the evolution of the MZR with formation
redshift seen here is inconsistent with the observations 
in dwarf galaxies. \citet{Kirby13} measured metallicities 
of dwarf galaxies based on measurements of individual stars. 
The authors established that a single MZR applies to all Local 
Group galaxies with $10^{3.5}<M_*/M_\odot<10^9$ regardless 
of their star formation histories. Dwarf irregular galaxies 
with gas present today have the same MZR as dwarf 
spheroidal satellites with no gas present today. If
this scenario held true in the more massive population, we would
expect a tight MZR regardless of formation redshift.
  
\subsection{On the slopes of the MZR relations}
\label{sec:slope}
Another interesting feature that emerges from the mock MZR is the
curve that bends toward lower metallicities at the low-mass end, even
though we constructed the mock MZR from a linear relation (black
dashed line in Figure \ref{fig:formation_mzr}b). The curve is similar
to the observed MZR. This suggests that the age--metallicity 
degeneracy causes the tendency to scatter toward
lower metallicity in low-mass galaxies. An explanation can be
found in the $\chi^2$ contours in Figure \ref{fig:wormplot}. At
[Fe/H]$\sim-0.2$, approximately where the change of slope occurs, the
contours of equal probabilities can be asymmetric, biasing toward
lower metallicities than the true values. In addition, lower
metallicities generally have larger uncertainties than the
uncertainties at solar metallicities, causing larger scatter at the
low-metallicity or the low-mass end. This finding could suggest that
the stellar MZR might in fact be a single power law with a similarly
tight dispersion, at least over the observed mass range of
$M_*\approx10^{9.7}M_\odot$ to $10^{11.5}M_\odot$. To confirm this,
individual spectra of low-mass galaxies with high S/N are required to
secure low uncertainties in [Fe/H], which is beyond our current work
but might be achieved by using gravitationally lensed galaxies, larger
telescopes, or longer exposure time.

The shallow slopes of $\sim0.16$ dex per $\log {\rm mass}$ found 
in both local and $z\sim0.4$ MZRs may give new insight 
into the strength of feedback in galaxies of the observed mass 
range. Strong star formation feedback generally results in 
a steep MZR \citep[e.g.][]{DeLucia12,Lu14}. \citet{Lu14} compared 
model predictions of both gas-phase and stellar MZRs 
from three independently developed semi-analytic models, 
namely the Croton model \citep{Croton06}, the Somerville model 
\citep{Somerville12,Porter14}, and the Lu model \citep{Lu11}. The 
authors found that the Croton model, which assumes a constant 
mass-loading factor, predicts the shallowest slope in the local 
stellar MZR, at $\sim 0.17$ dex per $\log {\rm mass}$
over the mass range of $10^8-10^{11} M_\odot$. In contrast, the
Lu model, in which the mass-loading factor is a strong function of halo
circular velocity, predicts a very steep MZR, at $\sim 0.6$ dex per
$\log {\rm mass}$ over the same mass range. The slope of the MZR in our
observations is consistent with the slope predicted from the Croton
model.

Remarkably, our results agree with what \citet{Lu14} found based on
the gas-phase MZRs at $z\lesssim1$. Among the three models considered,
the Croton model also describes the observations of gas-phase MZRs the
best. This might suggest that, over the observed mass and redshift
range, the amount of galaxy outflow is mainly a function of SFR and
does not have a strong additional dependence on galaxy mass. This
picture is closely related to the results from a simple closed box
model and from the FIRE hydrodynamical simulations, where stellar
metallicity is a strong function of gas fraction within a galaxy halo
\citep{Ma16}.

However, the slope of the stellar MZR found in this work is not
consistent with the slope of the stellar-MZR found in dwarf
galaxies. Dwarf galaxy satellites of the Milky Way exhibit a stellar MZR that is
consistent with an unbroken power law
\citep[e.g.][]{Grebel03,Kirby13}. Based on measurements of
metallicities of individual stars in local-group dwarf galaxies with
masses ranging from $M_*\sim 10^3$ to $10^{9} M_\odot$,
\citet{Kirby13} measured the slope of the MZR to be $0.30\pm0.02$ dex
in [Fe/H] per $\log {\rm mass}$. The slope of the dwarf galaxy MZR is
significantly larger than the slope found in this work, which is
based on individual integrated spectra of both local and $z\sim0.4$
galaxies with stellar masses $\gtrsim10^{9.5} M_\odot$. If both numbers
are correct, there must be a change of slope around the transition
mass.

The slopes of our observed stellar MZRs do not change 
significantly when we limit the sample to those with lower masses, 
$M_*<10^{10.5}M_\odot$. The slopes at this lower-mass end are 
$0.15\pm0.04$ and $0.33\pm0.16$ dex per log mass for $z\sim0$ and
$z\sim0.4$ sample respectively. Although the slope at the low-mass end 
of the $z\sim0.4$ sample may seem to suggest a change in slope, it is 
still consistent with the slope we found for the whole sample in Section 
\ref{sec:cl_results}. Furthermore, based on what we found from the 
mock data (Figure \ref{fig:formation_mzr}b), the age--metallicity degeneracy 
can cause the metallicity at the lower-mass end to bias low and create 
a seemingly steeper of slope.

Although \citet{Ma16} fit a single power law to the MZRs of
simulated galaxies over an entire 8 dex in mass, their FIRE
hydrodynamical simulations indeed seem to show a change of slope
around $M_*\sim10^{8.5}M_\odot$ for the MZR of $z\sim1.4-4$ simulated
galaxies. Dwarf galaxies exhibit a steeper slope than higher-mass
galaxies (as shown in \citeauthor{Ma16}'s Figure 4). Unfortunately, the sample sizes of
the simulated galaxies at $z=0$ and $z=0.8$ are also not large enough to
exhibit a clear change of slopes.

As discussed earlier, the slopes of the MZRs reflect the strength of
mass loading factors. If the change of slope is real, the mass at the
change of slope can suggest a mass below which feedback starts to have
additional dependence on other parameters. In fact, \citet{Lu17}
argued that at the low-mass regime, two different feedback mechanisms,
i.e., ejective and preventive feedback, are needed to explain both
the observed mass--metallicity relation and the stellar mass function.

 %%%%%%%%%%%%%%%%%%%%%%%%%%%%%%%%%%%%%%%%%%%%%%%%%%%%%%%%%%%%
%%%%%%%% SECTION 6 Environments %%%%%%%%%%%%%%%%%%%%%%%%%%%%%%%%%%%%%%%
%%%%%%%%%%%%%%%%%%%%%%%%%%%%%%%%%%%%%%%%%%%%%%%%%%%%%%%%%%%%% 
\section{Effect of Galaxy Environment}
\label{sec:environment}
Observing galaxies in galaxy clusters have a benefit of 
being able to obtain multiple spectra in a few telescope pointings. 
However, we have so far 
ignored the fact that our samples at $z\sim0.4$ are in
a dense cluster environment and treated them as if they were general
early-type galaxies. Here we discuss the impact of environment on our
results.

Not all properties of galaxies have been shown to correlate with
environment. Group and cluster environments show a higher fraction of
passive galaxies than that of the field environment
\citep[e.g.][]{Gerke07,Muzzin12,Koyama13}. At $z=0.4$, based on the
Hyper Suprime-Cam survey, the red fraction in cluster environments
(number of members $> 25$) is about 40\% higher in
$M_*\sim10^{9.5}M_\odot$ galaxies and about 20\% higher in
$M_*\sim10^{10.75}M_\odot$ galaxies \citep{Jian17}. 

However, neither galaxy size nor the galaxy stellar mass function (GSMF) seems to 
depend on global environment predominantly. \citet{Morishita17} found no significant 
differences in half-light radii between cluster or field systems in the 
\textit{Hubble Frontier Fields}. The shapes of the GSMFs of the general 
field and clusters are also mostly indistinguishable. The main difference 
is among the galaxies with $M_*\gtrsim10^{11}M_\odot$, which are 
more enhanced in high-density environments \citep[e.g.][]{Calvi13,Malavasi17, Etherington17}. 
Therefore, the completeness of our sample relative to the cluster 
population should be more or less transferable to the completeness 
of the general population.

In terms of chemical composition, gas-phase metallicities
in star-forming galaxies have been shown to have slight or 
no correlation with environment. More metal-rich
galaxies, on average, reside in over-dense regions 
\citep[e.g.][]{Cooper08,Wu17}. \citet{Cooper08}
used strong emission lines to measure gas-phase metallicities 
of SDSS star-forming galaxies. The authors found that the 
offset in metallicity relative to the median gas-phase 
MZR as a function of galaxy overdensity is significant. 
However, the metallicity offset between the least dense and
densest environment considered in that study is less than 0.03
dex. For higher redshift galaxies, \citet{Kacprzak15} studied gas-phase 
metallicities of star-forming galaxies in a $z\sim2$ galaxy cluster. They  
found no distinguishable difference between the gas-phase MZR 
of field and cluster galaxies to within 0.02 dex.

The effects of environment on age and metallicity 
are also likely to be minimal for early-type galaxies.
In particular, \citet{Fitzpatrick15} found that SDSS quiescent
early-type galaxies have slight variations in age with 
environment. Isolated galaxies have the youngest ages; 
brightest cluster galaxies are 0.02 dex older; and
satellites are 0.04 dex older than the isolated 
galaxies. There is no significant variation in Fe 
enrichment. Furthermore, \citet{Harrison11} measured 
ages and metallicities of early-type galaxies residing in 
four local galaxy clusters and their surroundings
extending to 10 $R_\text{vir}$. The ages and metallicities
were measured via spectrophotometric indices. 
They found no dependence of age or metallicity on the 
locations of galaxies in the clusters, i.e., those in the 
clusters or in the clusters' outskirts. \citeauthor{Harrison11} 
concluded that galaxy mass plays a major role in 
determining stellar populations.

In conclusion, cluster environment can affect the chemical abundance 
in galaxies. However, the effect seems to be small and weaker in 
stellar metallicities than in gas-phase metallicities. 

%%%%%%%% SECTION 7 Conclusion %%%%%%%%%%%%%%%%%%%
\section{Summary}
In this paper, we measured ages and metallicities of 62 individual
quiescent galaxies in the $z\sim0.4$ galaxy cluster Cl0024+17. The
quiescent galaxies were selected based on the EW of the [\ion{O}{2}]
$\lambda3727$\angstrom\ emission line and FUV$-$V color. The final sample spans the
stellar mass range from $10^{9.7}$ to $10^{11.5}M_\odot$ with
$\sim50\%$ completeness for $M_*\gtrsim10^{9.7}M_\odot$. We employed a
full spectrum fitting technique by adopting FSPS models
\citep{Conroy09} and the assumption of single stellar populations
(SSPs). We examined the accuracy of our fitting technique in several
aspects, including varying the signal-to-noise ratios, testing the
validity of the SSP assumption, and comparing with previous
measurements in the literature. Our age and metallicity measurements have
typical uncertainties of $<0.15$ dex. We also measured ages and
metallicities from a subsample of local SDSS quiescent galaxies from
\citet{Gallazzi05} and constructed the MZR of the local quiescent
galaxies based on our measurements. We used this local MZR to compare
with the MZR of $z\sim0.4$ galaxies. We find the following:
\begin{enumerate}
	\item We considered three functions (logarithmic, quadratic, 
	and linear) to fit the MZRs at both redshifts. We found that 
	the linear function fits the observed MZRs the best. 
         \item We detect an evolution of the stellar MZR with observed 
         redshift at $>5\sigma$. The evolution is $0.16\pm0.03$ 
         dex from $z\sim0.4$ to $z\sim0$ or $0.037\pm0.007$ dex per Gyr. 
         The observed evolution is greater but consistent within $1\sigma$ and 
         $2\sigma$ uncertainties with predictions from the FIRE and 
         the EAGLE hydrodynamical simulations. Our results may have 
         emphasized the importance of recycling processes in hydrodynamical 
         simulations.
	\item The intrinsic scatter of the MZR is smaller than that 
	of the gas-phase MZR but comparable to the scatter in the 
	fundamental metallicity relation. The intrinsic scatter can be 
	explained by an evolution of the MZR with galaxy age or 
	formation redshift. The MZR of galaxies that formed earlier 
	offsets toward lower metallicity in the same manner as the 
	evolution of the gas-phase MZR\@. The offsets are significant 
	and not caused by the age--metallicity degeneracy. The evolution 
	of the MZR with formation time is $0.055\pm0.006$ dex 
	per Gyr, which is stronger than and can explain the evolution 
	with observed redshift.
	\item Based on constructing a mock MZR from a linear 
	relation, the age--metallicity degeneracy can cause the 
	MZR at the low-metallicity end to offset to lower metallicity, 
	creating a downward curve sometimes seen in measurements 
	of the MZR\@.
	\item The slope of the MZR is $\sim0.16\pm0.03$ dex per 
	$\log {\rm mass}$. The slopes are consistent with the predicted 
	slope from a semi-analytic model \citep[the Croton model,][]{Croton06} 
	in \citet{Lu14}, which employs a modest, galaxy mass-independent 
	mass-loading factor. Our results suggest that galaxy feedback 
	(in terms of mass-loading factor) might not have a strong 
	additional dependence on galaxy mass over the observed 
	mass and redshift range.
\end{enumerate}
We will investigate the evolution of the MZR further in our future
work using galaxies at higher observed redshifts. We
will also measure $\alpha$ enhancements, as indicators of the
star formation timescales of those population.
 
\section{Acknowledgements}
The authors thank the referee for a constructive and helpful report. 
We also thank Anna Gallazzi and Jieun Choi for kindly providing
data, catalogs, and feedback to the paper. Thank Xiangcheng Ma, 
Yu Lu, Shea Garrison-Kimmel and Robyn Sanderson for 
useful discussions. E.N.K.\@ and N.L.\@ acknowledge support from
the National Science Foundation grant AST-1614081.

\appendix
\section{Comparisons with existing measurements}
\label{sec:compare_measurement}
\begin{figure}
	\centering
	\includegraphics[width=0.45\textwidth]{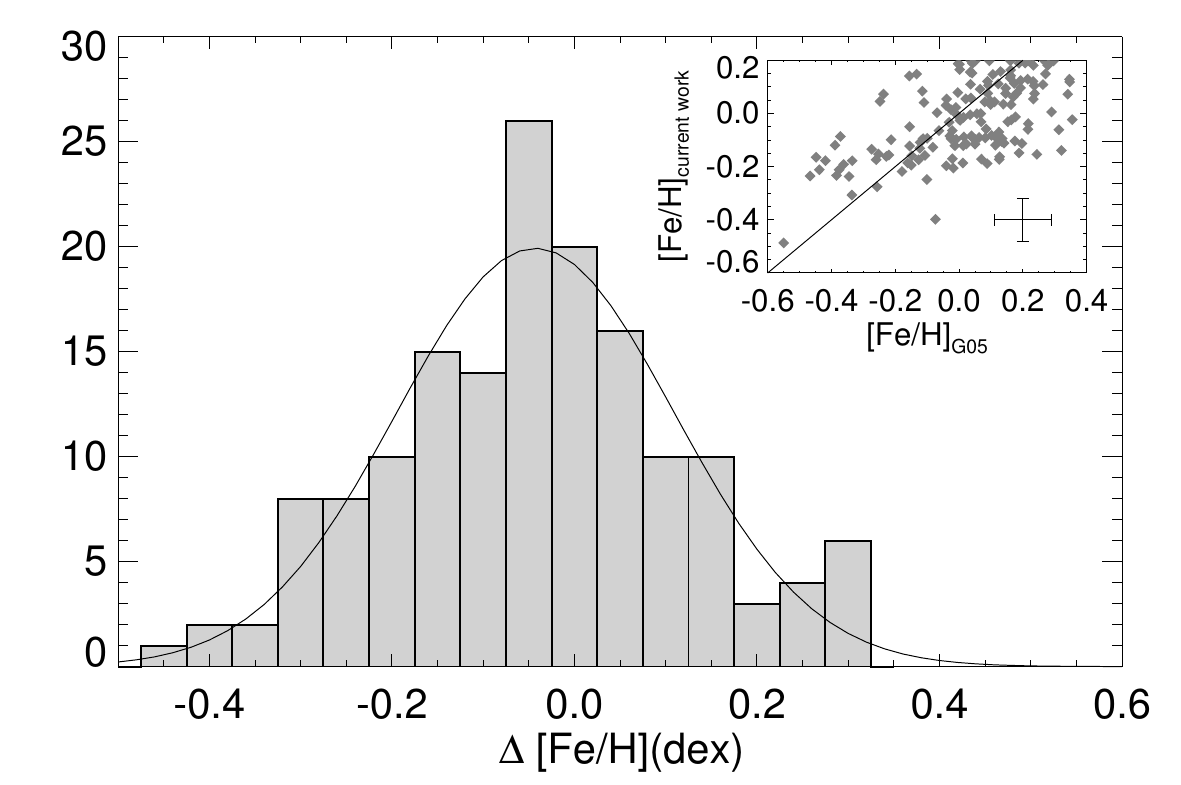}
	\includegraphics[width=0.45\textwidth]{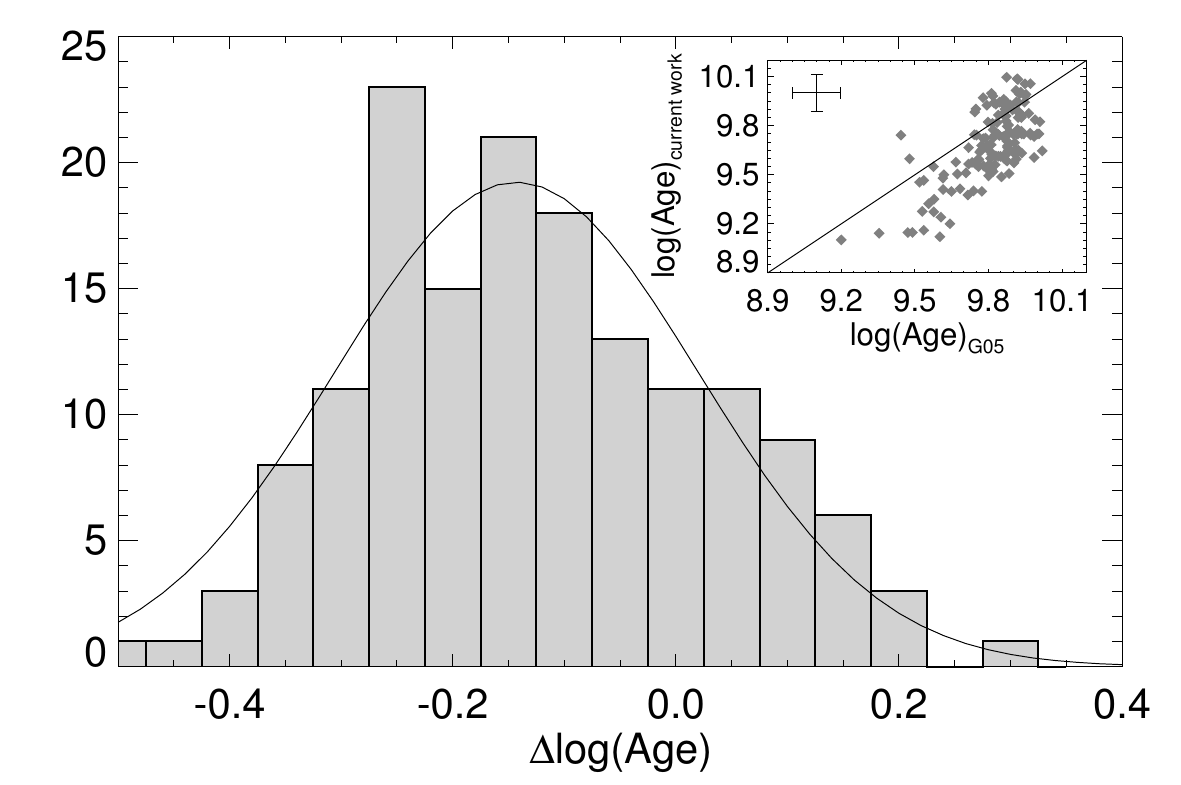}
	\caption{Comparison between the age and metallicities
measurement in this work and \citet{Gallazzi05} for the same
subsample of SDSS quiescent galaxies.}
	\label{fig:compare_sdss}
\end{figure}

We compare our measurements with literature measurements 
of the same galaxies \citep{Gallazzi05,Choi14}. We selected 
a subsample from 44,254 SDSS spectra from \citet{Gallazzi05}
that is comparable to our $z=0.4$ sample based on emission line
EWs and U$-$B colors. Since the SDSS spectra cover up to at least 8000
\angstrom\ in the rest frame but do not necessarily include the
[\ion{O}{2}]$\lambda3727$ \angstrom\ emission lines, we instead used the
criterion of rest-frame H$\alpha$ EW $<1$ \angstrom\ to define quiescent
galaxies. The limit of H$\alpha$ EW $=1$ \angstrom\ was chosen so
that the SFR is comparable to the SFR when the [\ion{O}{2}] EW is equal to
$5$ \angstrom. These limits were based on the SFR calibrations from
\citet{Kewley04} assuming color $B-V= 2$, a typical color limit for
quiescent galaxies \citep[e.g.][]{Schawinski14} and no dust
extinction. The color cut of U$-$B $>1$ is to make sure that the 
contamination from star-forming galaxies is minimized \citep{Mendez11} in a similar manner to 
the color cut in our $z\sim0.4$ sample. We selected all but at most 8 random quiescent early-type
galaxies from each bin of 0.1 dex in logarithmic mass spanning the stellar mass range
from $10^9$ to $10^{11.5} M_\odot$. This sums to a subsample of 155
quiescent galaxies. To be consistent with the observed $z\sim0.4$
spectra, we limited the wavelength range of the SDSS spectra to
3700--5500 \angstrom. We repeated the age and metallicity measurements
in the same manner as in Section \ref{sec:model}.

Our measured metallicities agree reasonably well with
the values measured by \citet{Gallazzi05}. The results are shown in Figure
\ref{fig:compare_sdss}. The differences in the [Fe/H] measurements
follow a Gaussian distribution with a width of $\sim 0.15$ dex,
peaking at $-0.04$ dex. Though the width of the metallicity differences
is comparable to the typical measurement uncertainty of 0.12 dex quoted by
\citet{Gallazzi05}, there are some systematic differences in the
measured [Fe/H]. The corner plot in Figure \ref{fig:compare_sdss}
shows that the metallicities we measured are slightly lower than those measured by
\citet{Gallazzi05} at high metallicities but the offsets
reverse at lower metallicities. We argue that the main reasons
for the discrepancy are the differences in the stellar libraries used
in generating model spectra, which will be discussed below together
with the uncertainties in age measurements.

The differences in age measurements show larger discrepancies than
those of metallicities. The distribution of differences in age peaks
at $\sim-0.15$ to $-0.25$ dex with a Gaussian width of 0.16 dex. The
galaxies were generally younger than reported by
\citet{Gallazzi05} by about 0.2 dex. The discrepancy is likely
SPS-model dependent. \citet{Gallazzi05} computed 5 spectrophotometric
indices from the BC03 stellar population synthesis, which is based on
the STELIB spectral library \citep{LeBorgne03}. The FSPS models used in the
current work are based on the MILES spectral library
\citep{Sanchez-Blazques06}. Although both spectral libraries are
empirical, the STELIB library contains fewer stars - 249 stellar
spectra as compared to 945 spectra in the MILES library. Very few
stars in the STELIB library are at non-solar metallicities \citep{Conroy10}.

\citet{Koleva08} compared three spectral synthesis models of single
stellar populations. In particular, the authors inverted the
parameters from the SSP spectra produced by BC03 using a grid of
models made with the Vazdekis/MILES \citep{Vazdekis10} and the
Pegase-HR spectral synthesis model \citep{LeBorgne04}. The two models
are based on the MILES library and the ELODIE library
\citep{Prugniel07}, respectively. They found that at approximately
solar metallicity, the ages retrieved by both Vazdekis/MILES and
Pegase-HR are $\sim0.2$ dex younger than the input ages in BC03 when
the input age is greater than $\sim5$ Gyr or $\log({\rm Age/yr}) \gtrsim 9.7$
\citep[see][Figure 2a]{Koleva08}. The trend reverses at $\sim0.2-0.3$
dex above the solar metallicity or 0.5 dex below the solar
metallicity. Our finding is somewhat consistent with this result,
in which we measured the ages to be $\sim0.15-0.25$ dex younger 
than measured by \citet{Gallazzi05}, as shown in Figure \ref{fig:compare_sdss}.

\begin{figure*}
	\centering
	\includegraphics[width=0.75\textwidth]{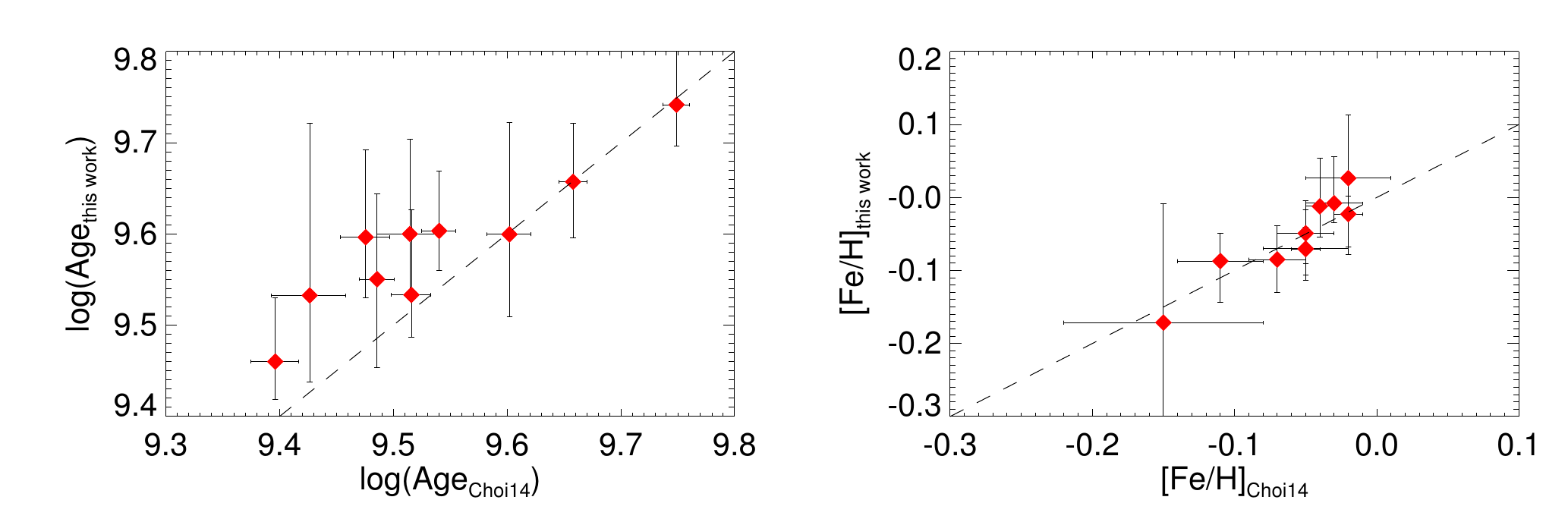}
	\caption{Comparison between the age and metallicities
measurement in this work and \citet{Choi14} for the same set of
stacked spectra.}
	\label{fig:compare_choi}
\end{figure*}

Lastly, we found good agreement between our measurements and those by
\citet{Choi14} in both ages and metallicities. We obtained a set of
stacked spectra compiled by \citet{Choi14}\footnote{Kindly obtained via private
communication.}. These spectra were stacked from individual spectra of
galaxies observed by the AGN and Galaxy Evolution Survey
\citep[AGES;][]{Kochanek12}, in bins of redshifts from $z=0.3$ to 0.7
and masses from $M_*=10^{10.2}$ to $10^{11.3} M_\odot$. The spectral
resolution is 6 \angstrom, roughly double the resolution of our DEIMOS
data. The wavelength coverage of the AGES spectra is 4000 to 5500
\angstrom, which does not cover the full age-sensitive Balmer break
region.

We found that our measurements of [Fe/H] are consistent with the
values measured by \citet{Choi14} (see Figure
\ref{fig:compare_choi}). For ages, all the values are consistent
within 0.1 dex. However, there are higher
discrepancies in the populations younger than 3.5 Gyr old, in the
sense that we measured the ages to be slightly older than
reported by \citet{Choi14}. The trend of this discrepancy is opposite
to the discrepancy found when we compared our age measurements with
\citet{Gallazzi05}.

Discrepancies between our work and \citet{Choi14} likely 
arise from differing wavelength coverage and the spectral models 
used. Because the AGES spectra only cover from 4000 to 5000 
\angstrom, higher order Balmer lines, which contain age 
information, are not present in the spectra. Moreover, \citet{Choi14} 
used the SPS model from \citet[][CVD12]{ConroyVandokkum12}. 
The main difference between the FSPS and the CVD12 model 
is that the latter allows abundances to be at non-solar ratios. 
Therefore, \citet{Choi14} fit the spectra for abundances of 
individual elements including [Mg/Fe], [O/Fe], [C/Fe], [N/Fe], 
etc. Though it would be beneficial to measure individual elements 
in our observed spectra, using the solar metallicity models without adjusting
individual elements can still provide a reasonable fit to the spectrum
\citep{ConroyVandokkum12}. Besides, the CVD12 is not yet readily
applicable to our data at lower masses because the model is limited to
fairly a small range around solar metallicity ([Fe/H]$\in(-0.4,0.4)$) and age greater than 3
Gyr. In fact, the lower age limit at 3 Gyr in the CVD12 might be
responsible for the small age discrepancies found in Figure
\ref{fig:compare_choi}.

\begin{longrotatetable}
\begin{deluxetable*}{lllrllrllllrllr}
\tablewidth{0pt}
\tabletypesize{\scriptsize} 
\tablecolumns{15}
\tablecaption{Catalog of Measured Age sand Metallicities\label{tab:catalog}}
\tablehead{\colhead{No.}&\colhead{RA} & \colhead{DEC} & \colhead{$\log M_*$} & \colhead{[Fe/H]} & \colhead{Age} & \colhead{S/N} & \colhead{} & \colhead{No.}&\colhead{RA} & \colhead{DEC} & \colhead{$\log M_*$} & \colhead{[Fe/H]} & \colhead{Age} & \colhead{S/N} }
\startdata
  1 & 00 25 51.07 & +17 08 42.4 &  10.8 &  $+0.10^{+0.08}_{-0.10}$ &    $  2.4^{+0.3}_{-0.2}$ &  16.0&   & 32 & 00 26 31.73 & +17 12 24.1 &  11.1 &  $-0.09^{+0.12}_{-0.07}$ &    $  3.3^{+1.1}_{-0.8}$ &  15.4 \\
  2 & 00 25 54.52 & +17 16 26.4 &  10.4 &  $-0.16^{+0.19}_{-0.17}$ &    $  2.9^{+2.3}_{-1.0}$ &   9.0&   & 33 & 00 26 31.81 & +17 11 56.9 &  10.4 &  $-0.02^{+0.08}_{-0.08}$ &    $  7.1^{+1.7}_{-1.0}$ &   8.0 \\
  3 & 00 25 57.73 & +17 08 01.5 &  10.4 &  $-0.12^{+0.08}_{-0.09}$ &    $  5.7^{+1.5}_{-2.0}$ &   8.4&   & 34 & 00 26 32.50 & +17 10 26.0 &  10.9 &  $-0.07^{+0.12}_{-0.20}$ &    $  2.5^{+0.8}_{-0.3}$ &  10.6 \\
  4 & 00 26 04.44 & +17 20 00.6 &  10.5 &  $-0.22^{+0.39}_{-0.17}$ &    $  2.0^{+0.7}_{-0.8}$ &   8.6&   & 35 & 00 26 32.71 & +17 07 56.0 &  10.3 &  $-0.20^{+0.13}_{-0.27}$ &    $  2.0^{+0.4}_{-0.4}$ &  10.0 \\
  5 & 00 26 05.80 & +17 19 19.0 &  10.9 &  $-0.05^{+0.07}_{-0.10}$ &    $  5.4^{+2.1}_{-2.6}$ &  11.9&   & 36 & 00 26 33.54 & +17 09 23.9 &  10.6 &  $-0.11^{+0.14}_{-0.60}$ &    $  1.3^{+0.7}_{-0.1}$ &  19.3 \\
  6 & 00 26 05.84 & +17 19 19.0 &  10.9 &  $-0.23^{+0.18}_{-0.20}$ &    $  6.0^{+4.3}_{-2.2}$ &  11.5&   & 37 & 00 26 33.60 & +17 09 20.2 &  10.5 &  $-0.13^{+0.09}_{-0.07}$ &    $ 10.0^{+1.4}_{-2.2}$ &   8.1 \\
  7 & 00 26 06.95 & +17 19 42.8 &  10.8 &  $-0.14^{+0.06}_{-0.05}$ &    $  4.6^{+0.7}_{-1.0}$ &  25.7&   & 38 & 00 26 33.66 & +17 09 31.0 &  10.5 &  $-0.07^{+0.07}_{-0.07}$ &    $  4.7^{+1.1}_{-1.7}$ &  12.3 \\
  8 & 00 26 08.85 & +17 09 54.7 &  10.7 &  $-0.14^{+0.12}_{-0.14}$ &    $  2.9^{+1.0}_{-0.4}$ &  18.3&   & 39 & 00 26 33.81 & +17 12 16.6 &  10.6 &  $-0.02^{+0.15}_{-0.07}$ &    $  3.4^{+1.5}_{-0.6}$ &   8.4 \\
  9 & 00 26 09.66 & +17 11 13.5 &  10.4 &  $-0.42^{+0.15}_{-0.15}$ &    $  7.5^{+1.9}_{-3.2}$ &   8.1&   & 40 & 00 26 34.35 & +17 10 22.1 &  11.1 &  $-0.16^{+0.07}_{-0.07}$ &    $  4.1^{+0.7}_{-1.0}$ &  27.4 \\
 10 & 00 26 13.92 & +17 13 34.9 &  10.1 &  $-0.33^{+0.24}_{-0.17}$ &    $  2.4^{+0.7}_{-0.4}$ &   8.1&   & 41 & 00 26 34.59 & +17 10 16.4 &  10.8 &  $-0.25^{+0.10}_{-0.15}$ &    $  5.6^{+2.4}_{-1.0}$ &  17.9 \\
 11 & 00 26 15.16 & +17 18 15.6 &  10.4 &  $-0.08^{+0.16}_{-0.31}$ &    $  2.7^{+2.2}_{-0.5}$ &  10.8&   & 42 & 00 26 34.98 & +17 10 21.3 &  10.5 &  $+0.04^{+0.09}_{-0.06}$ &    $  4.0^{+2.3}_{-1.7}$ &  16.4 \\
 12 & 00 26 18.45 & +17 07 01.1 &  10.3 &  $-0.27^{+0.18}_{-0.12}$ &    $  4.0^{+3.0}_{-1.0}$ &   9.0&   & 43 & 00 26 35.70 & +17 09 43.1 &  11.5 &  $+0.01^{+0.06}_{-0.03}$ &    $  5.6^{+1.0}_{-0.6}$ &  27.8 \\
 13 & 00 26 21.49 & +17 14 11.8 &  10.6 &  $-0.08^{+0.06}_{-0.08}$ &    $  5.5^{+1.5}_{-2.0}$ &  13.3&   & 44 & 00 26 36.77 & +17 09 28.7 &  10.1 &  $-0.23^{+0.12}_{-0.15}$ &    $  4.0^{+2.3}_{-1.1}$ &  10.5 \\
 14 & 00 26 22.90 & +17 12 31.4 &  11.1 &  $+0.20^{+0.04}_{-0.09}$ &    $  2.8^{+0.2}_{-0.1}$ &  23.7&   & 45 & 00 26 37.27 & +17 10 00.2 &  10.6 &  $+0.10^{+0.10}_{-0.12}$ &    $  5.6^{+3.1}_{-1.7}$ &  15.3 \\
 15 & 00 26 22.91 & +17 12 31.3 &  11.1 &  $-0.05^{+0.10}_{-0.08}$ &    $  4.3^{+1.6}_{-2.0}$ &   9.9&   & 46 & 00 26 37.52 & +17 09 08.7 &  10.6 &  $-0.18^{+0.07}_{-0.07}$ &    $  6.9^{+1.0}_{-0.9}$ &  16.8 \\
 16 & 00 26 24.82 & +17 12 21.5 &  10.9 &  $-0.14^{+0.11}_{-0.30}$ &    $  5.2^{+5.2}_{-2.0}$ &  12.3&   & 47 & 00 26 37.90 & +17 09 22.0 &  11.2 &  $-0.01^{+0.05}_{-0.06}$ &    $  4.8^{+1.5}_{-0.8}$ &  26.3 \\
 17 & 00 26 26.11 & +17 11 57.8 &  10.3 &  $-0.02^{+0.15}_{-0.15}$ &    $  5.4^{+3.2}_{-3.6}$ &   8.3&   & 48 & 00 26 37.91 & +17 09 37.8 &  10.7 &  $-0.07^{+0.09}_{-0.05}$ &    $  3.8^{+1.2}_{-0.8}$ &  14.5 \\
 18 & 00 26 27.12 & +17 12 25.9 &  10.9 &  $-0.16^{+0.12}_{-0.15}$ &    $  9.0^{+2.7}_{-3.1}$ &  12.1&   & 49 & 00 26 38.41 & +17 09 58.7 &  10.3 &  $-0.14^{+0.13}_{-0.09}$ &    $  3.2^{+1.1}_{-0.6}$ &  10.7 \\
 19 & 00 26 27.98 & +17 11 37.7 &  10.5 &  $-0.30^{+0.12}_{-0.22}$ &    $  8.5^{+3.1}_{-2.5}$ &   8.2&   & 50 & 00 26 38.65 & +17 09 14.5 &   9.7 &  $-0.15^{+0.17}_{-0.28}$ &    $  3.9^{+4.4}_{-2.9}$ &  10.9 \\
 20 & 00 26 29.11 & +17 10 24.7 &  10.7 &  $-0.07^{+0.07}_{-0.04}$ &    $  3.7^{+1.0}_{-0.7}$ &  18.0&   & 51 & 00 26 38.80 & +17 09 59.5 &  10.7 &  $-0.11^{+0.21}_{-0.18}$ &    $  2.7^{+2.3}_{-1.0}$ &  24.5 \\
 21 & 00 26 29.50 & +17 10 32.6 &  10.5 &  $-0.15^{+0.12}_{-0.14}$ &    $  3.1^{+1.2}_{-0.5}$ &  18.3&   & 52 & 00 26 40.13 & +17 08 21.5 &   9.9 &  $-0.60^{+0.08}_{-0.09}$ &    $  4.6^{+0.7}_{-0.9}$ &   8.7 \\
 22 & 00 26 29.92 & +17 10 06.8 &  10.3 &  $-0.21^{+0.04}_{-0.21}$ &    $  1.3^{+0.0}_{-0.1}$ &  16.4&   & 53 & 00 26 41.16 & +17 10 01.8 &  11.1 &  $+0.01^{+0.19}_{-0.25}$ &    $  3.0^{+2.7}_{-1.4}$ &  16.0 \\
 23 & 00 26 30.08 & +17 07 49.1 &   9.8 &  $-0.21^{+0.29}_{-0.26}$ &    $  2.1^{+0.8}_{-0.5}$ &   7.9&   & 54 & 00 26 43.23 & +17 08 41.0 &  10.1 &  $-0.25^{+0.36}_{-0.18}$ &    $  4.0^{+2.2}_{-5.1}$ &  11.5 \\
 24 & 00 26 30.76 & +17 12 26.3 &  10.3 &  $-0.15^{+0.13}_{-0.22}$ &    $  1.6^{+0.6}_{-0.3}$ &  13.2&   & 55 & 00 26 43.70 & +17 07 12.8 &  10.0 &  $-0.06^{+0.15}_{-0.17}$ &    $  3.0^{+1.5}_{-0.4}$ &   8.2 \\
 25 & 00 26 31.00 & +17 17 09.0 &  10.0 &  $+0.10^{+0.06}_{-0.10}$ &    $  2.3^{+0.3}_{-0.2}$ &  17.7&   & 56 & 00 26 44.68 & +17 08 33.8 &   9.9 &  $-0.79^{+0.35}_{-0.13}$ &    $  6.8^{+1.6}_{-0.9}$ &   8.3 \\
 26 & 00 26 31.04 & +17 11 09.3 &  10.4 &  $-0.09^{+0.08}_{-0.22}$ &    $  4.0^{+2.7}_{-1.7}$ &  12.5&   & 57 & 00 26 48.22 & +17 10 46.3 &  10.2 &  $-0.27^{+0.06}_{-0.16}$ &    $  1.7^{+0.2}_{-0.2}$ &   8.3 \\
 27 & 00 26 31.22 & +17 12 08.4 &  10.7 &  $+0.06^{+0.13}_{-0.08}$ &    $  3.1^{+1.7}_{-1.1}$ &  21.3&   & 58 & 00 26 48.30 & +17 12 35.4 &  10.6 &  $-0.28^{+0.05}_{-0.18}$ &    $  1.3^{+0.0}_{-0.1}$ &  16.2 \\
 28 & 00 26 31.40 & +17 17 00.3 &  11.1 &  $-0.06^{+0.14}_{-0.17}$ &    $  5.0^{+2.5}_{-3.3}$ &  13.5&   & 59 & 00 26 51.84 & +17 08 39.9 &  10.5 &  $-0.21^{+0.10}_{-0.20}$ &    $  6.5^{+2.8}_{-1.6}$ &  15.7 \\
 29 & 00 26 31.41 & +17 10 55.6 &  11.3 &  $-0.15^{+0.04}_{-0.06}$ &    $  6.3^{+0.7}_{-0.8}$ &  20.9&   & 60 & 00 26 54.38 & +17 08 27.0 &  11.0 &  $+0.10^{+0.11}_{-0.09}$ &    $  3.5^{+2.8}_{-1.4}$ &  24.8 \\
 30 & 00 26 31.41 & +17 10 27.1 &  10.0 &  $-0.27^{+0.13}_{-0.16}$ &    $  5.0^{+1.4}_{-3.5}$ &   8.0&   & 61 & 00 27 07.76 & +17 10 49.5 &  11.0 &  $+0.10^{+0.10}_{-0.13}$ &    $  3.0^{+1.0}_{-0.3}$ &  24.8 \\
 31 & 00 26 31.56 & +17 17 12.8 &  10.5 &  $-0.12^{+0.04}_{-0.06}$ &    $  5.9^{+0.7}_{-1.3}$ &  22.5&   & 62 & 00 27 25.16 & +17 07 22.4 &  10.1 &  $-0.51^{+0.24}_{-0.18}$ &    $  6.1^{+1.7}_{-3.4}$ &  12.0 \\
\enddata 
\tablecomments{The units of $M_*$, age, and S/N are $M_\odot$, Gyr, and \angstrom$^{-1}$, respectively. Full table is available online.} 
\end{deluxetable*}
\end{longrotatetable}

\end{document}